\documentclass[fleqn,usenatbib]{mnras}

\usepackage{amssymb}

\usepackage{newtxtext,newtxmath}
\usepackage{stfloats}
\usepackage[flushleft]{threeparttable}
\usepackage{subfigure}

\usepackage[T1]{fontenc}
\usepackage{setspace}

\usepackage{graphicx}	
\usepackage{amsmath}	

\usepackage{amssymb}	
\usepackage[justification=centering]{caption}
\usepackage{float}
\usepackage{longtable}
\usepackage{supertabular}

\title[Periodic X-ray Sources in GCs]{A Chandra Survey of Milky Way Globular Clusters -- IV. Periodic X-ray sources}

\author[Bao, Li, Cheng, Belloni]{
Tong Bao$^{1,2}$\thanks{E-mail: baotong@smail.nju.edu.cn}, 
Zhiyuan Li$^{1,2}$\thanks{E-mail: lizy@nju.edu.cn}, Zhongqun Cheng$^{3,4}$,
Diogo Belloni$^{5}$
\\
$^{1}$School of Astronomy and Space Science, Nanjing University, Nanjing 210046, China\\
$^{2}$Key Laboratory of Modern Astronomy and Astrophysics (Nanjing University), Ministry of Education, Nanjing 210046, China\\
$^{3}$School of Physics and Technology, Wuhan University, Wuhan 430072, China\\
$^{4}$WHU–NAOC Joint Center for Astronomy, Wuhan University, Wuhan 430072, China\\
$^{5}$Departamento de Física, Universidad Técnica Federico Santa María, Av. España 1680, Valparaíso, Chile
}

\date{Accepted XXX. Received YYY; in original form ZZZ}

\pubyear{2015}

\begin{document}
\label{firstpage}
\pagerange{\pageref{firstpage}--\pageref{lastpage}}
\maketitle

\begin{abstract}
We present a systematic search for periodic X-ray sources in 10 Galactic globular clusters (GCs) utilizing deep archival {\it Chandra} observations. By applying the Gregory-Loredo algorithm, we detect 28 periodic signals among 27 independent X-ray sources in 6 GCs, which include 21 newly discovered ones in the X-ray band. The remaining 4 GCs exhibit no periodic X-ray sources, mainly due to a relatively lower sensitivity of the data. 
Through analysis of their X-ray timing and spectral properties, complemented with available optical and ultraviolet information, we identify 21 of these periodic sources as cataclysmic variables (CVs). Combining with 11 periodic CVs in 47 Tuc similarly identified in the X-ray band, we compile the most comprehensive sample to date of GC CVs with a probable orbital period.
The scarcity of old, short-period CVs in GCs compared to the Galactic inner bulge and solar neighborhood, can be attributed to both a selection effect favouring younger, dynamically-formed systems and the hindrance of CV formation through primordial binary evolution by stellar dynamical interactions common to the GC environment.
Additionally, we identify a significant fraction of the GC CVs, most with an orbital period below or within the CV period gap, as probable magnetic CVs, but in the meantime there is a deficiency of luminous intermediate polars in the GC sample compared to the solar neighborhood.

\end{abstract}

\begin{keywords}
X-ray: binaries -- (Galaxy:) globular clusters -- stars: kinematics and dynamics -- novae, cataclysmic variables

\end{keywords}



\section{Introduction}

Globular clusters (GCs) are ancient, gravitationally bound systems characterized by a high stellar density and frequent stellar dynamical interactions. 
As such, GCs have been widely recognized as factories and reservoirs of close binaries, in particular low-mass X-ray binaries (LMXBs) harboring an accreting black hole (BH) or neutron star (NS), cataclysmic variables (CVs) harboring an accreting white dwarf (WD), coronally active binaries (ABs), as well as their potential descendants, e.g. millisecond pulsars (MSPs) and blue straggler stars. Many of these exotic objects are also expected to become detectable gravitational wave (GW) sources.
In the dense GC environment, the formation and evolution of close binaries are strongly affected by, or even directly produced in, stellar dynamical encounters (including flybys, tidal captures, and collisions). Moreover, in the course of such encounters, close binaries can release (or sometimes absorb) a substantial amount of kinetic energy \citep{1975MNRAS.173..729H,1975AJ.....80..809H}, thereby playing a crucial role in the gravothermal evolution of the host cluster \citep{2003gmbp.book.....H}. 
A deep understanding of close binaries in GCs, especially their formation channels and demography, is thus an important part of our general understanding of not only the profound astrophysics of binary formation and evolution but also the long-term evolution of GCs and their potential role as factories of GW sources and massive BH seeds.

Compared to normal stars, close binary systems can exhibit a significantly higher luminosity at certain wavelengths, making them more readily distinguishable even inside the dense cluster cores. 
As such, close binaries can serve as a unique tool for investigating the fundamental dynamical processes directly related to their own formation and evolution. 
In the X-ray band, while the prevalence of LMXBs as luminous X-ray sources (with a luminosity $L_{\rm X} \gtrsim 10^{35}~\text{erg~s}^{-1}$) in GCs was already established back in the {\it Uhuru} era \citep{1975ApJ...199L.143C,1975Natur.253..698K}, it is the advent of the {\it Chandra X-ray observatory} that enabled the routine detection of weak X-ray sources ($10^{30}~{\rm erg~s}^{-1}\lesssim L_{\rm X} \lesssim 10 ^{34}~{\rm erg~s}^{-1}$), which encompass CVs, ABs, quiescent LMXBs (qLMXBs) and MSPs, in many Galactic GCs. 
This facilitated the acquisition of a large sample of weak X-ray sources, ranging from a few to several hundred sources per cluster, which proved invaluable for statistical analyses and population studies \citep{2002ApJ...569..405P,2005ApJ...625..796H,2012ApJ...756..147M}.

While it is generally accepted that LMXBs are  over-abundant in GCs (with respect to the field) as a result of efficient dynamical formation, it is far less clear whether the same is true for the weak X-ray sources, especially CVs. \citet{2003ApJ...591L.131P} and \citet{2006ApJ...646L.143P} found a positive correlation between the number of detected CVs and the stellar encounter rate in a small sample of GCs, which they suggested to be evidence for a dynamical origin of the majority of CVs in GCs, similarly to the well-established case of LMXBs. 
However, \citet{2018ApJ...858...33C} used the cumulative X-ray emissivity (with luminous LMXBs subtracted) as a proxy of the abundance of weak X-ray sources (mostly CVs and ABs) and found an under-abundance, rather than overabundance, in most GCs relative to the field (see also \citealp{2020MNRAS.492.5684H}).
\citet{2018ApJ...858...33C} and \citet{2018ApJ...869...52C} suggested that this can be understood as stellar encounters being efficient in disrupting a large fraction of primordial, wide binaries before they can otherwise evolve into CVs and ABs. 
On the theoretical side, the MOCCA simulations \citep{2016MNRAS.462.2950B} also predicted that detectable CVs in GCs are predominantly composed of CVs formed via the common envelope phase rather than dynamical interactions.
The most recent MOCCA simulations \citep{2019MNRAS.483..315B} supports this assertion, even though they also suggest that strong dynamical interactions play a sufficiently important role, being able to trigger CV formations in binaries that would not have naturally evolved into CVs.
In addition, for pre-CVs dynamically formed in the core, the simulations predict that they cannot stay in the core after their formation,
because they are likely to acquire a sufficiently large recoil velocity during the dynamical encounter and get expelled from the core.

This implies that the spatial (radial) distribution of the weak X-ray sources is promising for probing the role of the dynamical processes related to close binary formation and evolution. 
Recently, \citet{2019ApJ...876...59C,2019ApJ...883...90C,2020ApJ...892...16C,2020ApJ...904..198C} conducted a series of studies examining the radial surface density profile of weak X-ray sources (mostly CVs and ABs) in 47 Tuc, Terzan 5, M28 and $\omega$ Cen, which are among the few GCs with a sufficiently large number of detected X-ray sources for a statistical analysis of this kind.
Their findings provide clear evidence for mass segregation. Specifically, the X-ray sources that are close binaries have a higher average mass compared to single stars and are more prone to sinking into the cluster core through two-body relaxation. 
In addition, despite the tendency to overlook the outer regions of GCs in both observational and theoretical investigations, \citet{2019MNRAS.483..315B} studied the CV populations in this region and found results aligned closely with those obtained by \citet{1997MNRAS.288..117D}. 
These authors discovered that a considerable proportion of pre-CVs may originate far from the central regions of GCs, where the detrimental effects of a crowded environment are less pronounced. 
Over time, these pre-CVs evolve and eventually develop into CVs, particularly in less evolved clusters. 
Notably, \citet{2019MNRAS.483..315B} determined that, on average, a substantial fraction (around half) of the detectable CV population is expected to exist beyond the half-light radius. They suggested that future observations should also prioritize the search for CVs in regions located further away from the cluster centres.

Nevertheless, observing GC CVs poses challenges. Various methods have been traditionally employed to identifying CVs, such as far-ultraviolet (UV) imaging with the Hubble Space Telescope (HST) \citep{2018MNRAS.475.4841R}, H$\alpha$ imaging \citep{2010ApJ...722...20C} and strong variability during CV outbursts \citep{1996ApJ...471..804S}, all of which require high sensitivity and high resolution observations. 
The high angular resolution achieved with {\it Chandra} enables the detection of CV candidates in the X-ray band, especially in the GC cores. 
When combined with HST optical counterparts, many CV candidates have been identified \citep{2010ApJ...722...20C}. However, it is important to note that a large fraction of faint CVs ($L_{\rm X} \lesssim 10^{32}~\text{erg~s}^{-1}$) in many GCs lack secure optical counterparts.
Furthermore, given the crowded nature of GCs and the intrinsic faintness of the CV population, confirming CV candidates spectroscopically is almost always a hard task \citep{2019A&A...631A.118G}. 

Therefore, we propose the use of periodic X-ray variations as an effective {\it tracer} for close binaries.
Especially in the case of magnetic CVs like polars, they exhibit a distinctive ``two-pole'' behaviour in their light curves, with nearly half of the cycle displaying a valley of near-zero X-ray flux (e.g, AM Her, \citealp{1985A&A...148L..14H}). 
Highly inclined binaries would also exhibit an eclipsing behaviour in their light curves, which points to an orbital modulation. 
Besides, most CVs have an orbital period shorter than 8 hours to satisfy the condition of {\it Roche lobe} overflow \citep{2011ApJS..194...28K}.
This helps to reduce the contamination from other interacting binaries like accreting NS or BH from a low mass star, a significant fraction ($\sim 42\%$) of which can have a much longer orbital period 
 \citep{2023A&A...675A.199A}.
By utilizing this tracer and with complementary information from optical, UV and radio observations, one can identify CV candidates among the numerous weak X-ray sources within GCs.

From the CV perspective, the orbital period plays a pivotal role as an observational tool in the study of their formation and evolution. This period is closely linked to the response of the donor star to mass transfer within the system and is associated with distinct features predicted by the standard CV evolution model \citep{1995cvs..book.....W}. These include a period gap of 2--3 hours and a period minimum around 80 minutes, which result from angular momentum loss (AML) mechanisms shaping CV evolution.
CVs with orbital periods longer than roughly 3 hours primarily experience AML through magnetic braking, while gravitational radiation dominates in CVs with shorter periods (less than about 2 hours). The interplay between these mechanisms influences the angular momentum and orbital dynamics of CVs. CVs within the period gap exhibit reduced mass transfer rates and become much less luminous.
CVs in the solar neighborhood provides compelling evidence supporting the predicted periodic characteristics of the standard model of CV evolution \citep{2003A&A...404..301R,2023MNRAS.524.4867I}. By probing the orbital period distribution of a certain CV population, valuable insights can be gained regarding the underlying physical processes governing their formation and evolutionary paths.

The first statistical study of the GC CV orbital period relied on a very limited sample of 15 CVs with known orbital periods \citep{2012MmSAI..83..549K}. 
Most of these CVs were found in the massive cluster 47 Tucanae through dedicated HST observations \citep{2003ApJ...596.1177E,2003ApJ...596.1197E},  although three of them were subsequently denied \citep{2018MNRAS.475.4841R}. 
Recently, \citet{2023MNRAS.521.4257B} conducted a comprehensive study of periodic signals from X-ray sources in 47 Tuc, utilizing both Chandra and eROSITA observations, in conjunction with multi-wavelength data. 
They identified 11 CVs with periodic X-ray signals within the core of 47 Tuc. 
This sample, although still small in size and likely biased (see discussions in Section~\ref{sec:discuss}), exhibits a significantly different distribution of orbital periods compared to field CVs.
This underscores the strong potential of using periodic X-ray sources to probe the CV population, and more generally close binaries, in the unique GC environment.

A systematic search for periodic X-ray signals from Galactic GCs is the goal of the present work, which is also the fourth paper in a series of an archival Chandra survey of X-ray emission from Galactic GCs. In our previous works, we have studied the emissivity and abundance of weak X-ray sources \citep{2018ApJ...858...33C,2018ApJ...869...52C} and searched for the X-ray signal from a putative intermediate-mass black hole \citep{2022MNRAS.516.1788S} in about half of the known Galactic GCs. 

This paper is organized as follows. In Section \ref{sec:data}, we describe the Chandra observation of our targets and data reduction procedure. Section \ref{sec:timing} provides a brief introduction to the method and process of detecting X-ray periodic signals and reports the candidate periodic signals/sources.
The X-ray spectral properties of the periodic sources are presented in Section \ref{sec:spec}.
Section \ref{sec:class} 
details a tentative classification of the periodic sources based on their temporal, spectral, and multi-wavelength properties, if available.
In Section \ref{sec:discuss}, we discuss the formation and evolution of GC CV populations in the scope of the frequency of the detected periodic X-ray signals in each GC and their collective orbital distribution, which provide compelling evidence for a dynamical origin of a fraction of the GC CVs. 
Finally, a summary of our study is provided in Section \ref{sec:summary}.

\begin{table*}
\centering
\caption{Basic information of the GCs}
\label{tab:obsinfo}
\begin{threeparttable}
\begin{spacing}{1.5}
\setlength{\tabcolsep}{1.5mm}{
\begin{tabular}{lcccccccccc}
\hline
\hline
GC Name & RA & DEC & ObsID & Exposure & $D$ & $r_{\rm h}$ & $r_{\rm c}$ & $[\mathrm{Fe} / \mathrm{H}]$ & Mass & Number of sources\\
 & (deg) & (deg) & & (ks) & (kpc) & (arcmin) & (arcmin) & & ($10^5\rm~M_{\odot}$) & \\
 (1) & (2) & (3) & (4) & (5) & (6) & (7) & (8) & (9) & (10) & (11)\\
\hline
47 Tuc & 6.02363 & -72.08128 & 78, 953, 954, 955, & 535.1 & 4.5 & 3.17 & 0.36 & -0.76 & 11.8 & 537 (184)
\\ (NGC 104) & & & 956, 2735, 2736, & & & &
\\ & & & 2737, 2738, 3384, & & & &
\\ & & & 3385, 3386, 3387, & & & &
\\ & & & 15747, 15748, 16527, & & & &
\\ & & & 16528, 16529, 17420 & & & &
\\
\hline
$\omega$ Cen & 201.69683 & -47.47958 & 653, 1519, 13727, 13726 & 290.9 & 5.2 & 5.00 & 2.40 & -1.62 & 25.7 & 300 (86)
\\ (NGC 5139) & & & & & & & 
\\
\hline
NGC 6121 & 245.89675 & -26.52575 & 946, 7446, 7447 & 119.2 & 1.9 & 4.33 & 1.17 & -1.20 & 1.52 & 117 (29)
\\
\hline
M62 & 255.30333 & -30.11372 & 2677, 15761 & 144.4 & 6.4 & 0.92 & 0.22 & -1.29 & 9.50 & 146 (36)
\\
(NGC 6266) & & & & & & & 
\\
\hline
NGC 6304 & 258.63438 & -29.46203 & 8952, 11073 & 102.7 & 5.8 & 1.42 & 0.21 & -0.59 & 1.68 &  183 (24)
\\
\hline
NGC 6397 & 265.17538 & -53.67433 & 79, 2668, 2669 & 339.7 & 2.4 & 2.97 & 0.05 & -1.95 & 0.91 & 376 (129)
\\ & & & 7461, 7460 & & &
\\
\hline
Terzan 5 & 267.02042 & -24.77917 & 3798 10059 13225, & 745.6 & 5.5 & 0.72 & 0.21 & -0.23 & 20.0 & 489 (130)
\\ & & & 13252, 13705, 13706, & & & & 
\\ & & & 14339, 14475, 14476, & & & &
\\ & & & 14477, 14478, 14479, & & & &
\\ & & & 14625, 15615, 15750, & & & &
\\ & & & 16638, 17779, 18881, & & & &
\\    
\hline
M 28 & 276.13671 & -24.86978 & 2683, 2684, 2685, & 325.9 & 5.4 & 1.97 & 0.24 & -1.45 & 3.71 & 502 (139)
\\ (NGC 6626) & & & 9132, 9133, 16748, &
\\ & & & 16749, 16750 &
\\
\hline
NGC 6656 & 279.09975 & -23.90475 & 5437, 14609 & 100.7 & 3.2 & 3.36 & 1.34 & -1.64 & 5.08 & 138 (20)
\\
\hline
NGC 6752 & 287.71712 & -59.98456 & 948 6612 , & 344.4 & 4.3 & 1.91 & 0.17 & -1.56 & 2.50 & 244 (62)
\\ & & & 19014 20121 20122 20123 & & & &
\\
\hline
M30 & 325.09217  & -23.17986 & 2679, 20725, 18997, & 330.1 & 8.0 & 1.03 & 0.06 & -2.12 & 1.93 & 84 (20)
\\ (NGC 7099) & & & 20726, 20732, 20731, & &
\\ & & & 20792, 20795, 20796 & &
\\
\hline
\end{tabular}
}
\end{spacing}
\begin{tablenotes}
      \small
      \item
      Notes: 
(1) Name of GC;
(2)-(3) Right Ascension and Declination (J2000) of the GC center;
(4)-(5) Chandra observation ID, effective exposure time in units of ks;
(6) Distance of GC, in units of kpc;
(7)-(8) Half-light radius and core radius in units of arcmin;
(9) Metallicity [Fe/H];
(10) Total mass in units of $10^5~\rm M_{\odot}$.
(11) The number of detected X-ray sources in the 0.5-8 keV band. The number of sources classified as ``bright'', i.e., with detected count $> 100$, is given in the parenthesis. 
Parameters in (2)-(10) are taken from \citet[][2010 ed.]{1996AJ....112.1487H} and \citet{2018ApJ...858...33C}.
\end{tablenotes} 
\end{threeparttable}
\end{table*}

\section{Sample selection and data preparation}
\label{sec:data}
Our sample selection started from the catalog of \citet[][2010 ed.]{1996AJ....112.1487H}, 
which contains 157 GCs. 
Among them, 81 GCs were observed by Chandra at least once with the Advanced CCD Imaging Spectrometer (ACIS) as of May 2022.
While in principle all these GCs may have source exhibiting periodic X-ray signals, in practice, 
we only selected 11 clusters for which a total exposure of at least 100 ks is available, ensuring an effective search for the periodic signal. 
The number of individual observations for the same GC ranges from 2 (M62, NGC\,6304, NGC\,6656) to 19 (47 Tuc), and
the total exposure ranges from 100.7 ks (NGC\,6656) to 745.6 ks (Terzan 5).
The main characteristics of the 11 clusters are listed in Table \ref{tab:obsinfo}. 
Among them, 47 Tuc was the target of  \citet{2023MNRAS.521.4257B}, who reported the detection of 20 periodic signals from 18 independent X-ray sources.
We will therefore focus on the other ten clusters in the following timing analysis (Section~\ref{sec:timing}).

Our data preparation follows the procedures of \citet{2023MNRAS.521.4257B}, which are outlined below.
We downloaded and uniformly reprocessed the raw data to obtain the level-2 event file for each observation,  following the standard procedure\footnote{http://cxc.harvard.edu/ciao} and using CIAO v.4.14 and calibration data files v.4.9.4. 
We have examined the light curve of each observation to ensure that the instrument background was sufficiently quiescent. Essentially all the science exposures were used for the timing analysis to ensure a maximally uninterrupted light curve within each observation. We further applied the CIAO tool \emph{axbary} to correct the photon arrival time of each registered event to the Solar System barycenter (i.e., Temps Dynamique Barycentrique time).

For each GC, we applied astrometry alignment among the individual observations, by matching the centroids of commonly detected point sources, using the CIAO tool \emph{reproject\_aspect} and taking the observation with the longest exposure as the reference frame.
A stacked counts map over the energy range of 0.5--8 keV was then created for the use of point source detection. 
Exposure maps and point-spread-function (PSF) maps were also
produced, according to a given enclosed count fraction (ECF). A fiducial source spectrum was applied for the exposure maps and PSF maps, which is an absorbed bremsstrahlung with a plasma temperature of 10 keV and a hydrogen column density derived by converting the E(B-V) values from \citet[][2010 ed.]{1996AJ....112.1487H} through $N_{\rm H} = 0.58 \times E(B-V) \times 10 ^{22} \rm cm^2$ \citep{2018ApJ...858...33C}. 
The exposure maps of a given GC field were stacked in the same way as for the counts maps.
The PSF maps were similarly stacked, weighted by the corresponding exposure maps.

Point source detection was conducted following the procedures outlined in \citet{2019ApJ...876...59C}.  Briefly, the identification of distinct sources was carried out using the CIAO tool {\it wavdetect} applied to the stacked 0.5-8 keV counts map. The algorithm was fed with the stacked exposure map and the 50\% ECF PSF map, setting a false-positive probability threshold of $10^{-6}$. The centroid of each source, determined by {\it wavdetect}, was further refined using an iterative maximum-likelihood method that considers the counts detected within the 90\% enclosed counts radius (ECR).
For Terzan 5, M28, and $\omega$ Cen, we utilized the catalogs provided by \citet{2019ApJ...883...90C,2020ApJ...892...16C,2020ApJ...904..198C}, respectively. For the remaining GCs, we followed the aforementioned procedure to generate the source catalog, which was then used to search for periodic X-ray signals.

\section{Timing analysis}
\label{sec:timing}

\subsection{Approach and Implementation}
The Gregory-Loredo (GL)  algorithm \citep{1992ApJ...398..146G}, essentially a Bayesian-based, phase-folding method, is effective in identifying periodic signals from X-ray data, which is often characterized by a moderate number of photon events and/or an irregular observing cadence.
The algorithm has been successfully employed to analyze 1-Ms {\it Chandra} observations of a Galactic bulge field known as the Limiting Window \citep[LW,][]{2020MNRAS.498.3513B}, 7-Ms observations of the {\it Chandra} Deep Field South \citep{2022MNRAS.509.3504B}, as well as deep {\it Chandra} and {\it eROSITA} observations of 47 Tuc \citep{2023MNRAS.521.4257B}, to detect periodic X-ray sources. 
The readers are referred to these papers for details of the GL algorithm and the specific implementation for Chandra data, as well as potential caveats.

Here we follow the recipe outlined in \citet{2020MNRAS.498.3513B} to perform a systematic search for periodic signals from the X-ray sources detected in the ten GCs.
We adopt a probability threshold ($P_{\rm GL}$) of 90\% for selecting the tentative periods returned by the GL algorithm. 
By design, the GL algorithm folds the photon time series at trial frequencies (periods). The resolution and range of searched period is compromised between efficiency and computational power.
We restrict our search in several period ranges: (10, 100) (100, 3000), (3000, 10000) and (10000, 50000) sec, with a frequency resolution of $10^{-6}$, $10^{-7}$, $10^{-8}$ and $10^{-9}$ Hz, respectively. 
Since the GL algorithm only determines the most probable period within a given range, we conduct a second search after each tentative detection, but exclude a narrow interval around the initially identified period to prevent the risk of missing a possible second period within the same searching range.

The chosen period ranges are optimal for detecting most orbital and WD spin periods of CVs, but also cover LMXBs with a relatively short orbital period. CVs typically have an orbital period shorter than 10 hours in order to satisfy the Roche lobe filling condition, but in rare cases, in which an evolved donor is involved, the orbital period can be significantly longer. 
For example, the famous 47 Tuc CV, AKO 9, consists of an accreting WD and a subgiant donor and has a well-determined orbital period of $\sim$26 hours \citep{2003ApJ...599.1320K}. The characteristic orbital period of CVs is thus covered by the latter two searching ranges. The second shortest range serves to probe the spin period of WDs in intermediate polars (IPs), whereas 
the shortest range is intended to probe pulsating signals potentially emanating from either NSs or rapidly spinning WDs. 
Examples of such targets include CTCV J2056-3014 ($P_{\rm spin}\sim 29.6$ sec; \citealp{2020ApJ...898L..40L}) and V1460 Her ($P_{\rm spin}\sim 38.9$ sec; \citealp{2020MNRAS.499..149A}).

The adopted frequency resolution corresponds to a period resolution of 0.1 to several seconds for the two longest period ranges, which
in practice serves as a lower limit of the uncertainty in the determined (orbital) period.
The {\it standard model} for CV evolution predicts an orbital period derivative $\dot{P}$ of the order of $10^{-13}-10^{-11}\rm~s~s^{-1}$ \citep{2011ApJS..194...28K,2020MNRAS.492.3323S}. Recent observations suggest that $\dot{P}$ can reach up to $\sim 10^{-9}\rm~s~s^{-1}$ during nova eruptions \citep{2023arXiv230713804S}. Even considering such an extreme intrinsic period change over a temporal baseline of $\sim$10 years (i.e., not atypical of the Chandra observations studied here), the period shift still does not exceed the period searching resolution of the GL algorithm. Therefore, we conclude that the intrinsic orbital period change of CVs has little effect in our detection.

For each source, we extract the 0.5 -- 8 keV counts within its 90\% ECR in each individual observation to generate a time series. For a small fraction of sources found in the crowded cluster core, we use a 75\% or 50\% ECR to minimize overlapping among different sources. 
These time series are then utilized as the input to the GL algorithm.
Since the GL algorithm, like most other phase-folding methods, manipulates the photon arrival times to evaluate the probability of periodic variation against a constant model, the background level is absorbed into the presumed constant.
Nevertheless, we provide an estimation of the background level for each source by extracting counts from within a concentric annulus with inner-to-outer radii of 2--4 times the 90\% ECR, masking any pixel falling within two times the 90\% ECR of neighbouring sources.

\subsection{Candidate periodic signals}
\label{subsec:candidate} 
As noted in \citet{2023MNRAS.521.4257B}, spurious signals could be reported by the GL algorithm for one of the following reasons: 

(i) Detector dithering.
The {\it Chandra}/ACIS operates with a regular dithering pattern to distribute photons over more CCD pixels. This pattern has a period of 706.96 s in pitch and 999.96 s in yaw{\footnote{https://cxc.harvard.edu/ciao4.4/why/dither.html}}. Signals detected at these two periods or their harmonics to within 1\%, are regarded as artificial signals. In our analysis, we have found several such signals, which are all detected in sources located near CCD gaps or edges.

(ii) Harmonics and sub-harmonics.
There is a well-known ambiguity in distinguishing the true period and its harmonics and sub-harmonics, i.e., integer division or multiplication of the true period{\footnote{It is noteworthy that harmonics are conventionally defined in frequency space. For convenience and not losing clarity, here we use the term harmonics on the period space.}}. 
In principle, the light curve can tell the true period only when we fully understand the mechanism behind the periodic variability. In the optical light curves of binaries, it is frequently observed that they exhibit two peaks attributed to the secondary star within a single cycle \citep{2003ApJ...596.1197E}.
The X-ray emission, however, primarily originates from the inner accretion disk or magnetic poles. For a magnetic CV, if the two poles alternately drift across the front side of the WD (the so-called two-pole behaviour), the resultant light curve will have nearly half of the cycle showing a valley of near-zero hard X-ray flux. On the other hand, if the X-ray-emitting pole is always visible (the so-called one-pole behaviour), it would produce a roughly constant light curve, although under certain condition dips can be present due to obscuration by the accretion stream \citep{2001cvs..book.....H}. 
Hence it is highly unlikely to present two nearly identical dips or peaks when folding the X-ray light curve. 
With this caveat in mind, we assume that there is no sub-harmonic in the periodic sources and always take the lowest period as the true period. 

(iii) Aperiodic variation. As discussed in \citet{2020MNRAS.498.3513B}, an aperiodically variable light curve may fool the GL algorithm to report a false period. 
To address this issue, we examine both long-term (inter-observation) and short-term (intra-observation) variations.  
To identify long-term variations, we define for each source an inter-observation variability index, VI = $S_{\rm max} /S_{\rm min}$ , where $S_{\rm max}$ and $S_{\rm min}$ are the maximum and minimum photon flux among all the observations, respectively. For a candidate periodic source with a strong inter-observation variation (defined by VI $>$ 10), we repeat the period search in two subsets of the light curve: one covering only the high state (defined as the observation[s] with the highest photon flux) and the other excluding the high state.  
To identify short-term variations, we inspect the light curve of individual observations for each source with a candidate period, and subsequently repeat the period search after discarding any observation(s) in which  significant short-term flares are present.
Only when a periodic signal survives the above procedures it is considered a genuine signal.
Moreover, we follow the merit and method of \citet{2023MNRAS.521.4257B} to evaluate for each signal a false alarm probability (FAP) due to red noise (see Appendix \ref{sec:confidence} for details).

After the above filtering, we have detected 28 periodic signals in 27 X-ray sources, among which one source (in M28) exhibits dual periods.
The 27 sources are distributed in 6 GCs, including $\omega$ Cen (2), M62 (3), NGC\,6397 (3), Terzan 5 (10), M28 (6) and NGC\,6752 (3).
Basic information of these sources/signals, including source coordinates, projected distance from the cluster center, the identified period, the GL probability, number of source and background counts, and source classification (Section~\ref{sec:class}), are summarized in Table \ref{tab:srcinfo}. 
It is noteworthy that the vast majority of the signals have $P_{\rm GL}$ greater than 0.99, and the lowest value of $P_{\rm GL}$ is 0.9628, significantly higher than the adopted threshold of 0.9.
The phase-folded light curves of the 28 signals are shown in Figure~\ref{fig:pfold_lc1}.

No periodic X-ray signal is found in the remaining four GCs, M30, NGC 6121, NGC 6304, and NGC 6656, which deserves some remarks.
Overall, we consider this to be mainly due to a lack of sufficiently bright sources (hence a light curve of sufficiently high signal-to-noise ratio [S/N]), rather than an intrinsic paucity of periodic signals, in these four GCs.
Intuitively, the detection efficiency of the GL algorithm depends primarily on the number of source counts and the amplitude of the periodic variation.
\citet{2020MNRAS.498.3513B} used simulated sinusoidal light curves to show that the GL detection rate is generally $\gtrsim 0.2$ when the source counts $C \gtrsim 100$, exceeds $\sim$0.5 when $C \gtrsim 300$, and rises rapidly to $\gtrsim 0.9$ at high variation amplitudes ($\gtrsim 60\%$).  
The number of ``bright'' sources, i.e., those with $C > 100$, is given for each GC in Table~\ref{tab:obsinfo}. 
Indeed, this number is the lowest ($<$ 30) in the four GCs without any detected periodic signal.  
This scarcity of bright sources in these four GCs seems to be due to either a relatively short exposure, or a relatively small cluster mass, or both (Table \ref{tab:obsinfo}).
In the case of M30, which has a relatively long exposure, most observations were conducted in the sub-array mode, resulting in a small field-of-view.

The last aspect that deserves discussion is the uncertainty in the reported period. The GL algorithm calculates the probability of periodic variation by integrating the odds ratio over the range of period searching, ultimately identifying the frequency (period) with the highest odds ratio. In an ideal scenario, the uncertainty in determining the frequency of a periodic signal should be constrained by the resolution of the frequency sampling (from $10^{-6}$ to $10^{-9}$ Hz in our work).
However, in practice, when comparing X-ray detection results with those obtained from other wavelengths or algorithms, it is crucial to consider the systematic errors inherent in periodicity searches.
Specifically, when dealing with a source exhibiting intrinsic periodic variability, the uncertainty in determining the period from observed data must account for the presence of noise. 
\citet{2023MNRAS.521.4257B} conducted numerical simulations to estimate the uncertainty in the GL-reported period for different types of light curves. 
Their findings indicated that the magnitude of uncertainty depends on various factors, including the amplitude of variability, noise level, photon counts, and more. 
In the case of strong signals, the uncertainty can be as low as 0.01\%. Conversely, for signals with a high noise, lower variability amplitude, or limited photon counts, the uncertainty may increase to 1\% -- 2\%.

\begin{table*}
\centering
\caption{Basic information of the periodic X-ray sources in six GCs}\label{tab:srcinfo}
\begin{threeparttable}
\begin{spacing}{1.5}
\begin{tabular}{llllllllllll}
\hline
\hline
GC & Seq    & RA       & DEC   & $R$   & $P$ & $P_{\rm GL}$ & $C$ & $C_{\rm B}$  & Eclipse & Class & Note  
\\ 
  &  & deg & deg & arcsec & second &  & counts & counts &  &  
\\
(1) & (2) & (3) & (4) & (5) & (6) & (7) & (8) & (9) & (10) & (11) & (12)
\\
\hline
$\omega$ Cen & 1 & 201.67081 & -47.40050 & 291.35 & 5889.28 & 1.0000 & 166 & 10.3 & & CV & XMM 67, 51d
\\
& 2 & 201.61938	 & -47.44087 & 234.52 & 13513.51 & 0.9957 & 509 & 5.6 & Yes & CV & XMM 49, 41d
\\
\hline
M 62 & 1 & 255.29844	& -30.11131 & 17.54 & 7006.55 & 0.9925 & 262 & 2.6 & & CV & s28
\\
& 2 & 255.30231	& -30.11777 & 14.90 & 25250.01 & 0.9710 & 472 & 6.1 & & CV & s06
\\
& 3 & 255.30641	& -30.10888 & 19.90 & 29411.76 & 0.9954 & 243 & 4.4 & & CV & s32
\\
\hline
NGC 6397 & 1 & 265.30641	& -53.66355 & 72.60 & 20321.07 & 1.0000 & 2289 & 4.6 & Yes & CV & CV6
\\
& 2 & 265.28893	& -53.58670 & 397.84 & 34084.32 & 0.9976 & 2121 & 10.6 & Yes & CV &
\\
& 3 & 265.1734	& -53.67207 & 9.17 & 40711.64 & 1.0000 & 9781 & 19.6 & Yes & CV & CV1
\\
\hline
Terzan 5 & 1 & 267.07537 & -24.71579 & 293.96 & 4817.42 & 1.0000 & 177 & 40.2 & & CV &
\\
& 2 & 267.04564 & -24.82514 & 185.55 & 5025.13 & 1.0000 & 99 & 8.2 & & CV &
\\
& 3 & 267.06387 & -24.75265 & 171.49 & 5304.76 & 0.9951 &  79 & 10.3 & & CV &
\\
& 4 & 267.01811	 & -24.77687 & 10.43 & 7417.30 & 1.0000 & 1115 & 42.4 & & CV & CX17
\\ 
& 5 & 267.01844 & -24.77747 & 8.12 & 13882.46 & 0.9999 & 2892 & 49.2 & & CV & CX6
\\
& 6 & 267.01911 & -24.77852 & 4.08 & 15723.27 & 1.0000 & 2086 & 66.8 & & tMSP & CX1
\\ 
& 7 & 267.01833 & -24.78456 & 20.75 & 22732.44 & 0.9943 & 832 & 14.9 & & CV & CX8
\\
& 8 & 267.01111 & -24.76757 & 50.95 & 29877.50 & 0.9628 & 4022 & 16.1 & Yes & CV & CX5
\\ 
& 9 & 267.02103 & -24.77827 & 3.90 & 31486.15 & 0.9999 & 1887 & 50.9 & & MSP & Ter5 P, CX10
\\
& 10 & 267.01929 & -24.77946 & 3.33 & 44483.99 & 0.9999 & 1126 & 57.4 & Yes & MSP & Ter5-VLA38,  CX13
\\
\hline
M 28 & 1\dag & 276.08422 & -24.90590 & 213.15 & 751.13  & 0.9995  & 416 & 6.2 & & CV &
\\
& 2 & 276.18980	 & -24.74253 & 491.98 & 5434.78 & 0.9967 & 479 & 130.8 & & CV &
\\
& 3 & 276.18772	& -24.89096 & 183.43 & 5738.66 & 0.9600  & 156 & 7.0 & Yes & CV &
\\
& 4\dag & 276.08422 & -24.90590 & 213.15 & 13622.12 & 0.9984 & 416 & 6.2 &  & CV & 
\\
& 5 & 276.13164& -24.87155 & 153.76 & 37586.38 & 0.9999 & 2857 & 11.4 &  & MSP & MSP H
\\
& 6 & 276.13540	& -24.86894 & 5.84 & 39692.81 & 1.0000 & 9211 & 27.6 & & tMSP & MSP I
\\
& 7 & 276.15939 & -25.02704 & 569.27 & 47846.89 & 0.9948 & 995 & 103.5 & & CV & 
\\
\hline
NGC 6752 & 1 & 287.71425	& -59.98476 & 5.24 & 33003.30 & 1.0000 & 1673 & 10.0 & & CV & CX5
\\
& 2 & 287.71497	& -59.98379 & 4.75 & 40985.29 & 0.9974 & 1803 & 10.8 & & CV & CX4
\\
& 3 & 287.76237	& -59.99503 & 89.76 & 46296.30 & 0.9977 & 298 & 3.3 & & AB & CX8
\\
\hline
\end{tabular}
\end{spacing}
\begin{tablenotes}
      \small
      \item
      Notes: 
(1) Name of GC.
(2) Sequence number for each periodic signal. The signals from the same source are marked by \dag.
(3) and (4) Right Ascension and Declination (J2000) of the source centroid.
(5) The projected distance from the cluster centre. 
(6) The modulation period determined by the GL algorithm. 
(7) The GL probability.
(8) The number of total counts in the 0.5--8 keV band.
(9) The number of estimated background counts.
(10) The presence/absence of eclipsing behavior in the phase-folded light curve.
(11) Tentative source classification. `tMSP' stands for transitional MSP.
(12) IDs and possible MSP counterparts assigned in the literature: 
\citet{2003A&A...400..521G}, \citet{2009ApJ...697..224H}, $\omega$ Cen;
\citet{2020MNRAS.498..292O}, M\,62;
\citet{2001ApJ...563L..53G}, NGC 6397;
\citet{2006ApJ...651.1098H}, \citet{2021ApJ...912..124B} Terzan 5;
\citet{2011ApJ...730...81B}, M 28; 
\citet{2014MNRAS.441..757F}, NGC 6752;

\end{tablenotes} 
\end{threeparttable}
\end{table*}

\begin{figure*}
\centering
\begin{subfigure}{}
    \includegraphics[scale=0.9]{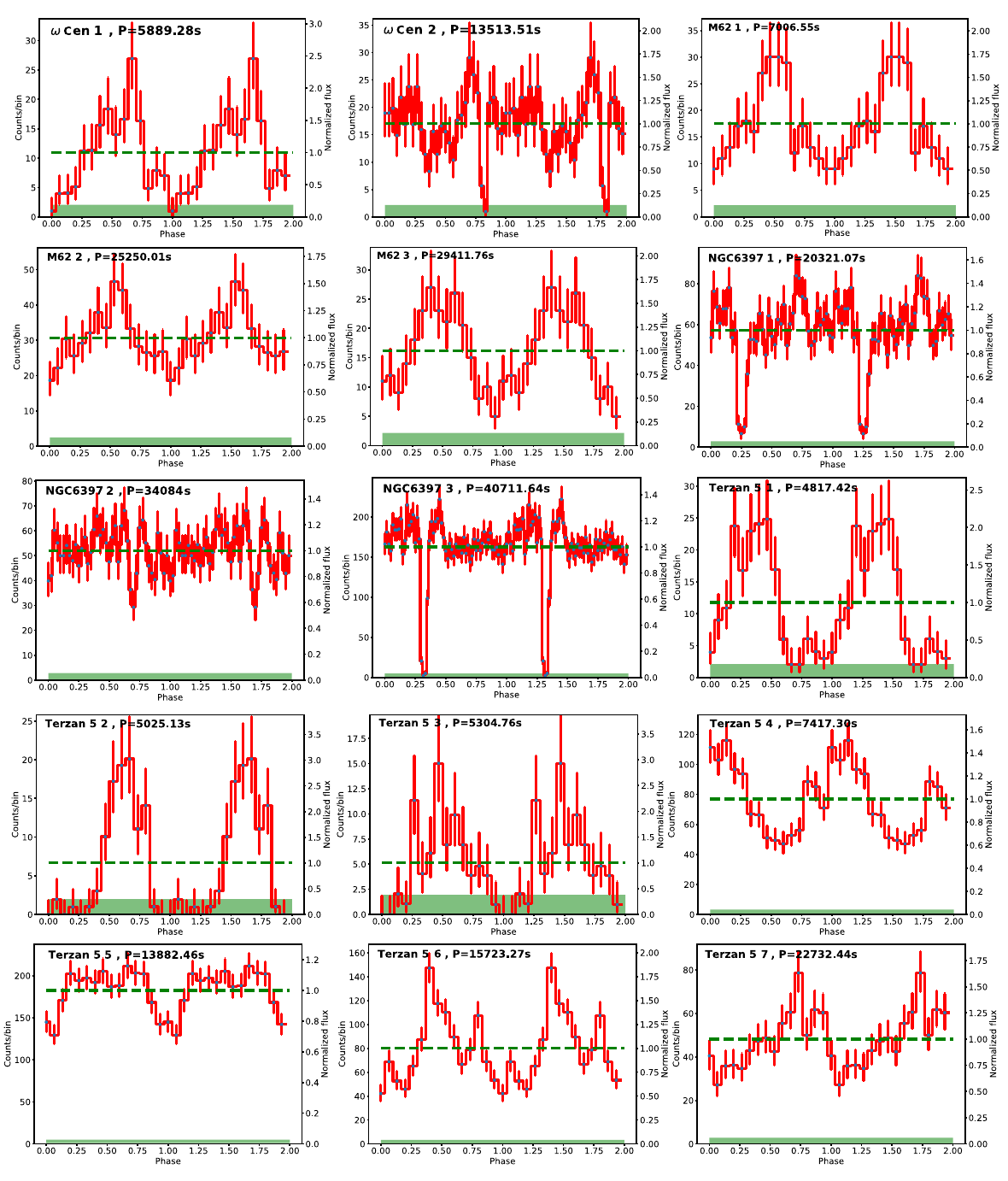}
    \caption{The 0.5--8 keV phase-folded light curve at the modulation period. The source name and the identified period are marked in each panel. The green dashed line represents the mean count rate, whereas the green strip marks the local background, the width of which represents 1\,$\sigma$ Poisson error.}
\label{fig:pfold_lc1}
\end{subfigure}
\end{figure*}

\begin{figure*}
 \ContinuedFloat
\centering
\begin{subfigure}{}
    \includegraphics[scale=0.9]{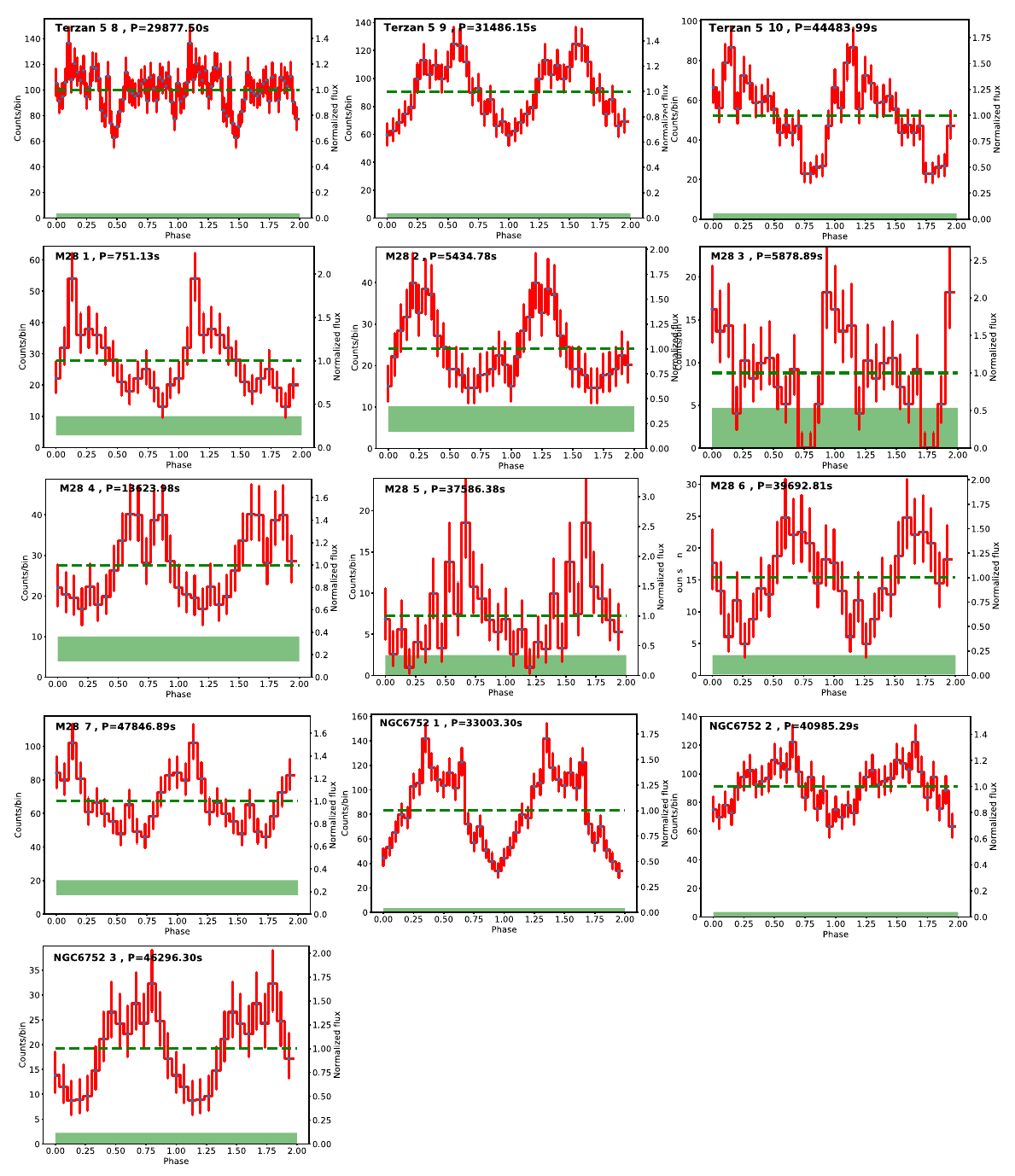}
    \caption{Continued}
\label{fig:pfold_lc2}
\end{subfigure}
\end{figure*}

\section{X-ray spectral properties}
\label{sec:spec}
Prior to classifying the identified periodic sources, it is crucial to examine their spectral properties, which shed light on their true nature. 
The source and background spectra are generated utilizing the CIAO tool \emph{specextract} from the aperture defined in Section \ref{sec:timing}. 
For a given source, spectra extracted from individual observations are co-added. 
For the two sources (Seq.6 of Terzan 5 and Seq.6 of M28) whose periodic signal is only detected in a high state, only observations during the high state are included.
We then apply an adaptive binning technique over 0.5-8 keV such that a minimum of 20 counts and a S/N greater than 2 per bin are achieved. XSPEC v12.13.0 are used for the spectral analysis \citep{1996ASPC..101...17A,2001ASPC..238..415D}.

We utilize a fiducial phenomenological spectral model, an absorbed bremsstrahlung continuum (\emph{tbabs*bremss} in XSPEC), to fit the spectra, which turns out to provide a reasonably good fit to all spectra.
It is expected that the periodic sources are mainly LMXBs, CVs, or ABs. 
In particular, the X-ray spectra of CVs are primarily originated from a collisionally ionized plasma and,  therefore, are expected to display metal emission lines \citep{2016ApJ...818..136X}. However, most of the 27 sources exhibit a featureless spectrum, which can be attributed to a moderate spectral S/N and a low metallicity characteristic of GCs.
Exception is found in three sources, Seq.2 of NGC\,6397, Seq.4 of Terzan 5 and Seq.7 of M28, which exhibit significant excess consistent with Fe lines at 6--7 keV. A Gaussian line is thus added to account for this excess. 
In several cases where a relatively hard spectrum is present, the bremsstrahlung temperature, $T_{\rm b}$, is not well constrained and is thus fixed at a fiducial value of 40 keV.
The unabsorbed 0.5--8 keV luminosity is reported, based on the fitted model and corrected for the spectral extraction aperture.

\begin{table}
\centering
\setlength\tabcolsep{3pt} 
\begin{threeparttable}
\caption{X-ray spectral properties of the periodic sources \label{tab:spec}}
\begin{spacing}{1.5}

\begin{tabular}{llllr}
\hline
\hline
GC Seq. & $N_{\rm H}$ & $T_{\rm b}$ & $\chi^2$(d.o.f) & $L_{\rm 0.5-8}$ \\
 & $10^{22}\rm~cm^{-2}$ & keV  & & $\rm 10^{31}~erg~s^{-1}$ \\
(1) & (2) & (3) & (4) & (5) \\
\hline
$\omega$ Cen & & & & 
\\
 1 & $0.6^{+0.5}_{-0.3}$ & $1.0^{+0.5}_{-0.4}$ & 0.71 (19)  & $5.5^{+0.8}_{-0.8}$
\\
 2 & $1.0^{+0.3}_{-0.3}$ & $6^{+5}_{-2}$ & 1.13 (56)  & $15^{+1}_{-1}$
\\
\hline
M 62 & & & & 
\\
 1 & $0.2^{+0.2}_{-0.1}$  & $16^{+89}_{-9}$  & 0.71 (29) &  $12^{+2}_{-2}$
\\
 2 & $0.10^{+0.10}_{-0.09}$ & $5^{+3}_{-1}$ & 0.92 (49) &  $14^{+1}_{-2}$
\\
 3 & $0.2^{+0.1}_{-0.1}$ & 40 (fixed) &  0.84 (30) &  $11^{+2}_{-2}$
\\
\hline
NGC 6397 & & & & 
\\
 1 & $1.0^{+0.1}_{-0.1}$ & $28^{+22}_{-10}$ & 0.85 (155) &  $11.0^{+0.4}_{-0.4}$
\\
 2 & $0.82^{+0.11}_{-0.09}$ & $12^{+8}_{-4}$ & 0.70 (146) &  $42^{+2}_{-2}$
\\
 3 & $0.73^{+0.04}_{-0.03}$ & $11^{+2}_{-2}$ & 1.12 (212)  & $38.5^{+0.7}_{-0.6}$
\\
\hline
Terzan 5 & & & &
\\
 1 & $2^{+1}_{-1}$ & $10^{+118}_{-5}$ & 0.58 (12) &  $6^{+1}_{-1}$
\\
 2 & $1.6^{+1.3}_{-0.7}$ & 40 (fixed)  & 1.62 (11) &  $4.0^{+0.8}_{-0.8}$
\\
 3 & $4^{+4}_{-2}$ & $3^{+30}_{-2}$ & 0.80 (7)  & $3.8^{+0.8}_{-0.8}$
\\
 4 & $3.2^{+0.4}_{-0.3}$ & 40 (fixed) & 0.87 (115) &  $20^{+1}_{-1}$
\\
 5 & $1.69^{+0.07}_{-0.06}$ & $7.8^{+1.1}_{-0.9}$ & 1.02 (158) &  $36^{+1}_{-1}$
\\
 $6^a$ & $1.7^{+0.1}_{-0.1}$ & 40 (fixed) & 0.88 (147)  & $27^{+1}_{-1}$
\\
 7 & $1.4^{+0.2}_{-0.2}$ & 40 (fixed) & 0.97 (87)  & $10.1^{+0.6}_{-0.6}$
\\
 8 & $1.8^{+0.1}_{-0.1}$ & $6.2^{+0.8}_{-0.7}$ & 1.05 (179)  & $57^{+1}_{-1}$
\\
 9 & $2.3^{+0.2}_{-0.2}$ & 40 (fixed) & 0.93 (150)  & $27^{+1}_{-1}$
\\
 10 & $2.2^{+0.3}_{-0.2}$ & 40 (fixed) & 0.80 (94)  & $15^{+1}_{-1}$
\\
\hline
M 28 & & & &
\\
 1\&4 & $5^{+1}_{-1}$ & 40 (fixed)  & 1.17 (45) &  $24^{+3}_{-3}$
\\
 2 & $0.5^{+0.2}_{-0.2}$ & $7^{+2}_{-1}$ &  1.17 (33)  & $9^{+1}_{-1}$
\\
 3 & $0.2^{+0.1}_{-0.1}$ & 40 (fixed) & 1.38 (19)  & $2.0^{+0.4}_{-0.4}$
\\
 5 & $0.1^{+0.2}_{-0.1}$ & 40 (fixed) &  1.08 (12) & $1.3^{+0.3}_{-0.3}$
\\
 $6^a$ & $0.22^{+0.02}_{-0.02}$ & $9^{+1}_{-1}$  &  1.02 (201) & $124^{+3}_{-3}$
\\
 7  & $<0.05$  & $1.01^{+0.07}_{-0.08}$ & 0.95 (46) &  $17 ^{+5}_{-4}$ 
\\
\hline
NGC 6752 & & & & 
\\
 1 & $0.22^{+0.05}_{-0.05}$ & $6^{+2}_{-1}$ & 1.12 (116) &  $14.0^{+0.6}_{-0.6}$
\\
 2 & $0.18^{+0.05}_{-0.04}$ & $4.7^{+1.0}_{-0.8}$ & 0.81 (118) &  $14.3^{+0.6}_{-0.6}$
\\
 3 & $0.3^{+0.2}_{-0.2}$ & $0.3^{+0.1}_{-0.1}$ & 1.4 (27)  & $4.6^{+0.5}_{-0.5}$
\\
\hline
\end{tabular}
\end{spacing}
\vskip-1.3cm
\begin{tablenotes}
      \small
      \item
      Notes: 
      (1) Source sequence number.
      $^a$The source is at a high state.
(2) Line-of-sight absorption column density
(3) The bremsstrahlung temperature, fixed at a value of 40 keV if the spectrum provides no strong constraint. 
(4) $\chi^2$ and degree of freedom of the best-fit model.
(5) 0.5--8 keV unabsorbed luminosity. 
Quoted errors are at the 90\% confidence level. 
\end{tablenotes} 
\end{threeparttable}
\end{table}

\section{Classifying the periodic sources}
\label{sec:class}
Once the periodic X-ray signals are identified, we attempt to classify the nature of the signal/source, i.e., whether the signal is modulated by orbital or spin motion, and whether the source is a CV, an LMXB, or something else.
As exemplified in \citet{2023MNRAS.521.4257B} for the periodic X-ray sources found in 47 Tuc, 
the X-ray temporal and spectral properties, when assisted with multi-wavelength data, can be very useful for the classification. 
However, it is notoriously challenging to distinguish among numerous UV or optical sources for the genuine counterpart of an X-ray source, particularly within the dense cluster core. 
Nevertheless, we consult available high-resolution HST observations and radio imaging, as well as the literature, to identify plausible counterparts on a best-effort basis.
For those periodic sources with absent multi-wavelength data, the tentative classification is based primarily on their X-ray temporal and spectral properties.
In what follows, we provide more details about the individual sources in each of the investigated GCs.

\subsection{$\omega$ Cen}
Two periodic sources are detected in this cluster. 
Despite not exhibiting a significant blue excess, the optical counterpart of Seq.1 (labeled \#67 by \citealp{2003A&A...400..521G}; 51d by \citealp{2009ApJ...697..224H}) displays strong H$\alpha$ emission \citep[who label it ``H$\alpha$-only'']{2013ApJ...763..126C}, which is typical of CVs or ABs.
However, the short period of only 1.6 hr makes it highly unlikely to be an AB.
Thus we classify this source as a CV and the period as an orbital period, which falls below the CV period gap. 
Seq. 2 of $\omega$ Cen (labeled \#49 by \citealp{2003A&A...400..521G}, 41d by \citealp{2009ApJ...697..224H}) is an eclipsing system with a period of 3.7 h (Figure~\ref{fig:pfold_lc1})
The blue $B-R$ colour and strong H$\alpha$ emission has led to the suggestion that this is a CV \citep{2013ApJ...763..126C}, which we follow here.

\subsection{M\,62}
Three periodic sources are detected in this cluster. The severe foreground extinction toward M62 precludes any UV information.
Nevertheless, all three sources were classified as a CV based on an X-ray color–luminosity diagram by \citet{2020MNRAS.498..292O}. 
We can classify Seq.1 as a polar based on its phase-folded light curve exhibiting a typical two-pole behavior, which can be understood as spin modulation on a magnetized WD synchronized with the binary orbit period \citep{2020MNRAS.498.3513B}.
For the other two sources, we tentatively classify them as CVs according to \citet{2020MNRAS.498..292O} and take the period as the orbital period, since both these two periods are probably too long to be the spin period of WDs.

\subsection{NGC\,6397}
Three periodic sources are detected in this cluster, and all three exhibit an eclipsing behavior (Figure~\ref{fig:pfold_lc1}), clearly pointing to an orbital period. 
A similar period was previously determined from ground-based optical time series for Seq. 1 (also known as CV 6) by \citet[0.2356 days]{2003AJ....125.2534K}, while the first eclipse was noted by \citet{2001ApJ...563L..53G}, who referenced a similar 11.3 hour period from HST imaging. 
Similarly, a corresponding period was previously identified for Seq. 3 (also known as CV 1) by \citet[0.472 days]{2003AJ....125.2534K}.

Both sources have been suggested to be CVs due to the presence of a bright UV counterpart \citep{2017MNRAS.469..267D}. 
On the other hand, Seq.2 is located outside the coverage of the optical/UV images. 
Nevertheless, we can reasonably classify it as a CV, due to the presence of a significant 6.7 keV line that is consistent with belonging to the Fe XXV K$\alpha$ complex, which is among the most commonly detected emission lines in CV spectra. The relatively high plasma temperature, $\sim$12 keV, disfavours the possibility of an AB. 

\subsection{Terzan\,5}
Terzan 5 is a heavily obscured (with an average color excess E(B-V) =2.38; \citealp{2012ApJ...755L..32M}) cluster located close to the Galactic center, which precludes UV information. 
It has the deepest Chandra data among all 10 GCs, from which 10 periodic sources are detected. 
Seq. 9 is known to be an MSP (Terzan 5 P, with a 0.3626 day orbital period; \citealp{2005Sci...307..892R,2017ApJ...845..148P}); the Chandra counterpart was identified by \citet{2021ApJ...912..124B}, who identified X-ray orbital modulation typical of redback pulsars (including an eclipse at orbital phase 0.25). 
Seq. 10 was inferred to be a likely redback MSP by \citet{2020ApJ...904..147U}, who found a radio VLA counterpart (Ter5-VLA38) and identified a 12.32-hour period from the Chandra data.
It is readily apparent that our period detection results show substantial discrepancies when compared to prior findings, by an amount of 118 seconds and 135 seconds, respectively.  Nevertheless, as we highlighted in Section \ref{subsec:candidate}, for sources with relatively modest variability amplitudes, the relative uncertainty in the period can rise to as much as 1\% or even 2\%. In the case of these specific sources, a difference exceeding 100 seconds with respect to previous results is entirely plausible.
Therefore, the GL algorithm confirms the periodicity in both sources, which can be taken as the orbital period. This is further supported by an eclipsing behavior in the case of Seq.10.

Seq.6 was suggested to be a possible candidate of a transitional millisecond pulsar (tMSP), given both the existence of a radio counterpart and the presence of a typical transitional state in the X-ray band \citep{2018ApJ...864...28B}. Its orbital modulation has been confirmed at 4.37 hr during its bright, accretion-powered state in 2003 and 2016.

Among the remaining periodic X-ray signals, Seq.1-Seq.3 all have a period shorter than 2 hr, i.e., falling below the orbital period gap of CVs. Their phase folded light curves share similar characteristics, with nearly half of the cycle exhibiting a high-amplitude peak and the remaining phase dominated by a constant valley. This so-called two-pole behavior serves as strong evidence for a magnetic CV origin. 
The identified period is thus taken as the orbital period for all these three sources.
Seq.4 exhibits a hard continuum and significant 6.7 and 7.0 keV lines, which also support the CV classification. Its period of 2.06 hr is taken as the orbital period, which notably falls on the lower edge of the period gap.

Seq.5, Seq.7 and Seq.8 exhibit longer periods. Due to the lack of an optical/UV/IR information, it is difficult to definitively classify these sources as either qLMXBs or CVs. 
However, we tentatively classify them as CVs based on their location (reasonably bright, with hard spectra) in the X-ray color-magnitude diagram \citep{2006ApJ...651.1098H,2006ApJ...646L.143P}.

\subsection{M\,28}
Six periodic sources are detected in this cluster, with the majority of them situated at a considerable distance from the center, beyond the coverage of the presently accessible HST images. 
The dual period of Seq.1 and Seq.4 originating from a single source is a strong evidence that the source is an IP, in which the shorter period (751 sec) represents the spin period, while the longer period (3.78 hr) corresponds to the orbital period.  
Seq 5 (designated as M\,28 H by \citealp{2011ApJ...730...81B}) and Seq.6 (known as M\,28 I by \citealp{2013Natur.501..517P}) are identified as radio pulsars. 
M\,28 H possesses a radio timing orbital period of 0.43502743 days, as cited by \citet{2010ApJ...725.1165P}. On the other hand, M\,28 I has an X-ray pulse timing orbital period of 11.025781 hours. 
Their respective X-ray orbital modulations were detected by \citet{2011ApJ...730...81B} for M\,28 H, and \citet{2022ApJ...941...76V} for M\,28 I.

We have independently verified these periods and included relevant details in Table \ref{tab:srcinfo}. 
Notably, the periodic signal of Seq.6 is exclusively detected during its accretion-powered state, particularly in the observations conducted in 2002 and 2015.
 
Regarding the remaining sources, the classification of Seq.2 as a CV is justified by the ``two pole'' like light curve, making it more likely to be a polar. Moreover, Seq.3 exhibits an eclipse, clearly pointing to an orbital period. 
A tentative classification of Seq.7 as a CV is based on its X-ray luminosity and thermal spectrum. Nevertheless, similar to the case of Terzan 5, the absence of any optical or UV counterpart prevents completely ruling out the possibility of Seq.7 being a qLMXB. 
Moreover, our confidence on the reality of this periodic signal is somewhat less than on the other sources, as indicated by the estimated FAP from our simulations (see Appendix \ref{sec:confidence}).

\subsection{NGC\,6752}
Three periodic sources are detected in this cluster. 
Among them, Seq.2 has long been recognized as a dwarf nova (\citealp{1996ApJ...473L..31B}, their star 1; the X-ray counterpart is CX 4 in \citealp{2002ApJ...569..405P}), with a reported optical period of 5.1 h only by folding the light curve. 

Analysis of the near-UV data suggested a period of 6.9 h, but the significance of this signal, based on the Lomb–Scargle method, was only marginal \citep{2012MNRAS.423.2901T}.   
Here we report a new X-ray period of approximately 11.4 hr with high confidence.
We suggest that the previously reported optical and UV periods could be just spurious. We take 11.4 hr as the orbital period of this source.

Seq. 1 was identified in the X-ray (as CX5) and optical by \citet{2002ApJ...569..405P}, who suggest it as an AB due to its H$\alpha$ excess and location, but note that its high X-ray/optical flux ratio suggests a CV. \citet{2012MNRAS.423.2901T} finds it to lie on the main sequence in a near-UV/U CMD, which might support an AB classification. \citet{2017ApJ...841...53L} argue that a CV nature is most likely based principally on the X-ray/optical flux ratio, while \citet{2021MNRAS.508.2823C} note its bizarre features (no blue colours in the UV colour-magnitude diagram, a small H$\alpha$ excess, and rapid X-ray variability), and conclude that it is most likely to contain a neutron star or black hole, either as a qLMXB or redback MSP.
Here we classify it as a CV although we cannot rule out the possibility of an accreting BH or a transitional MSP.

Seq. 3 was also identified in the X-ray (as CX8) by \citet{2002ApJ...569..405P}. \citet{2012MNRAS.423.2901T} suggest a faint blue source as the optical counterpart, and a CV nature. \citet{2017ApJ...841...53L} and \citet{2021MNRAS.508.2823C}  instead identify a pair of brighter, H$\alpha$-bright optical counterparts as foreground, chromospherically active M-dwarf binaries. 
The association of the X-ray source with these stars is strengthened by the X-ray spectrum, which is best fit by a double thermal plasma model (kT$\sim$0.2 and $\sim$1.7 keV), typical of chromospheric emission from ABs.
Regarding the previous classification based on the color-magnitude diagram, it should be noted that the UV emission from M dwarfs is predominantly chromospheric in nature. Therefore, the colors observed in this system are not unusual in the context of chromospheric emissions \citep{2013MNRAS.431.2063S,2020MNRAS.492.5684H}.

Moreover, the periodic signal observed in Seq.3, as well as in Seq.1, has a non-negligible FAP based on our simulations (see Appendix \ref{sec:confidence}).
However, the inter-observation light curves of Seq.1 strongly supports its intrinsic periodicity (see Appendix \ref{sec:longlc}).
Considering the above factors, we classify Seq.1 and Seq.2 as CVs and take the identified period as the orbital period. Finally, Seq.3 is classified as an AB in the foreground.

So far, we have discussed 28 periodic signals found in 27 sources distributed in 6 GCs.
We have classified 21 CVs, 5 MSPs and 1 AB (Table \ref{tab:srcinfo}) with a varied degree of confidence based on their X-ray timing and spectral properties, and multi-wavelength counterparts. 
Twenty-one of the periodic signals are newly discovered. 
In addition, 20 periodic signals have been identified in 18 X-ray sources in 47 Tuc, from the core region to the cluster outskirt, as presented in \citet{2023MNRAS.521.4257B}, 
which were classified into 11 CVs, 4 LMXBs, 2 ABs and 1 MSP. 

\section{Discussion} 
\label{sec:discuss}

Based on the classification outlined in Section \ref{sec:class}, our sample consists of 32 periodic CVs observed across 7 GCs, representing the most extensive sample of periodic CVs in GCs to date, approximately three times larger than the one compiled by \citet{2012MmSAI..83..549K}. 
These include 11 CVs in 47 Tuc, 2 in $\omega$ Cen, 3 in M 62, 3 in NGC\,6397, 7 in Terzan 5, 4 in M\,28, and 2 in NGC 6752.

\begin{figure*}
	\includegraphics[scale=0.56]{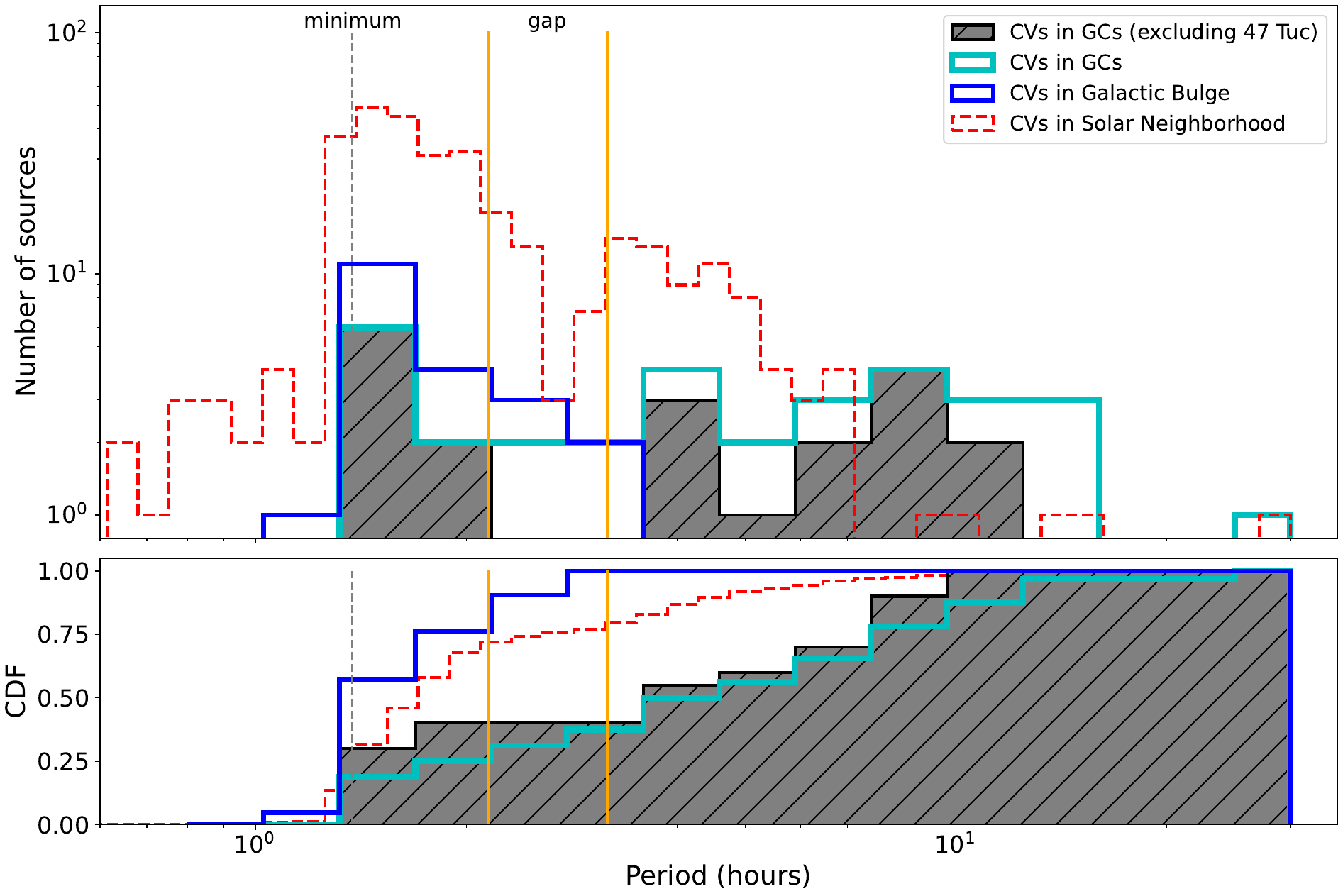}
    \caption{The orbital period distribution of the GC CVs (cyan histogram), in comparison with that of solar neighborhood CVs (red-dotted histogram) and Galactic bulge CVs (blue histogram). CVs in GCs excluding 47 Tuc are shown by the grey filled histogram. The period gap is delineated by a pair of vertical orange solid lines, and the period minimum by a vertical grey dashed line, the values of which are taken from \citet{2011ApJS..194...28K}. The bottom panel shows the cumulative distribution.}
    \label{fig:N_P}
\end{figure*}

We show in Figure \ref{fig:N_P} the orbital period distributions of the GC CVs together with the orbital period distribution of CVs in the solar neighbourhood based on the Sloan Digital Sky Survey \citep{2023MNRAS.524.4867I}, and the CVs in the Galactic bulge (i.e., the LW) CVs based on Chandra data  \citep{2020MNRAS.498.3513B}. 
Note that the solar neighborhood sample (326 CVs; represented by the red dashed histogram) is significantly larger than the bulge (20 CVs; blue solid histogram) and the GC (cyan histogram) samples. 

Two potential {\it caveats} related to the GC CV sample deserve remarks before further discussion. 
The first is the authenticity of the CV orbital periods. 
As explained in Section~\ref{sec:class} and \citet{2023MNRAS.521.4257B}, 13 out of the 32 ($\sim$ 41\%) GC CVs exhibit an eclipsing behavior, which clearly identifies an orbital period. 
Confidence with the remaining source varies. 
Five sources exhibit a ``two-pole'' characteristic in their phase-folded light curves, which suggests a good chance of being polars (and thus an orbital period equals the WD spin period, as in polars the entire binary is synchronized).
The IP identification for the two sources with dual periods also seems reasonable.
Fourteen out of the 32 sources have a known UV/optical counterpart or strong H$\alpha$ emission, which can be taken as good evidence for their CV nature.
Other three sources exhibit significant iron lines in their X-ray spectra, also a good indication of the CV nature.
Therefore, the genuineness of the identified orbital period, as well as the CV identification itself, seems to be secured for the majority of the periodic X-ray sources. 
The remaining sources, which accounts for only a small fraction of the total ($\sim19\%$, 6 of 32), lack additional supporting evidence.
Nevertheless, excluding these six sources would not affect the main conclusions drawn in this work.

The second caveat is the incompleteness and potential bias of the GC CV sample, which unfortunately are difficult to rigorously quantify.
In fact, even the much larger sample of the solar neighborhood CVs is subject to incompleteness and selection effects.
In general, the observed samples are biased towards relatively bright sources. 
Nevertheless, the solar neighborhood sample suffices to reveal the most fundamental properties of the CV period distribution, i.e., a period minimum at $\sim$ 80 minutes, a period gap between $\sim$ 2--3 hours, and a paucity of sources at periods longer than $\sim$10 hours (Figure \ref{fig:N_P}). 
As we argue below, the current sample of GC CVs, albeit biased and limited in size, can still uncover rich information about the formation and evolution of CVs in the exotic GC environment.

\subsection{CV population in individual GCs}
\label{subsec:CVpop}

\subsubsection{$\omega$\,Cen}
For the two periodic CVs in $\omega$ Cen, we suggest that they have formed through normal evolution of primordial binaries. 
This is due to the expectation that the relatively low stellar density of $\omega$ Cen leads to less frequent stellar encounters. Additionally, it has been suggested that numerous stellar mass BHs are present in $\omega$ Cen \citep{2019ApJ...877..122Y,2020ApJS..247...48K,2020ApJ...904..198C}, which, if true, should suppress CV formation via dynamical exchange. 
\citet{2009ApJ...697..224H}
estimated that at least 1/2 to 2/3 of the primordial binaries that would otherwise give rise to CVs are destroyed in $\omega$ Cen before they can evolve to that stage.
Therefore, detecting only two periodic CVs in such a massive GC with a considerable Chandra exposure appears reasonable.

\subsubsection{M\,62, NGC\,6397 and NGC\,6752}
M\,62, NGC\,6397 and NGC\,6752 are classified as core-collapsed GCs by \citet{1996AJ....112.1487H}. In such GCs, most BHs should have already been ejected after sinking into the GC center due to mass segregation \citep{1969ApJ...158L.139S}. Specifically, BHs rapidly mass-segregate towards the cluster core, forming a dense central BH sub-cluster. Within this BH-dominated core, three-body encounters generate numerous dynamically-hard BH binaries \citep{2015ApJ...800....9M}, which then interact with passing stars (including other BHs), transferring energy through scattering interactions. BHs undergoing these binary-mediated encounters receive significant recoil kicks that displace them away from the core \citep{1993Natur.364..421K,2021RNAAS...5...47R}.
Once all stellar-mass BHs have been dynamically ejected, the cluster core is no longer supported against collapse and lower-mass stellar populations (in particular WDs) efficiently segregate to the cluster center \citep{2021ApJ...917...28K}. 

At this point, pre-CVs and CVs can eventually segregate towards the central parts. Among the 8 periodic CVs detected in these three GCs, most of them (six out of eight) are found in the core region of their parent GCs. These CVs located in the central parts are consistent with being formed without any influence of dynamics. As suggested by \citet{2019MNRAS.483..315B}, non-dynamical pre-CVs are typically formed when the cluster is younger than $\sim$ 2 Gyr. Then, they spend another $\sim$ 10 Gyr as detached binary, having in turn enough time to eventually segregate towards the central parts depending on the energy source driving the GC evolution, until they finally become CVs, i.e. mass transfer starts. Some of them could have formed dynamically a few Gyr ago, eventually ejected from the central parts and migrated back to the central parts due to mass segregation.
It is noteworthy that one source, Seq.2 of NGC\,6397, is found at a large projected distance of $\sim 2.2~r_{\rm h}$, which might be due to an inefficient mass segregation there, but we cannot rule out the possibility that it is a foreground/background source, rather than a true member of the cluster. 

\subsubsection{Terzan\,5 and M\,28}
\label{subsubsec:outerCV}
Terzan 5 and M\,28 are currently known to host 39 and 13 MSPs, respectively \citep{2022ApJ...941...22M,2022ApJ...941...76V}. In general, MSPs are X-ray faint sources with $L_{\rm X}\lesssim 10 ^{31}~{\rm erg~s}^{-1}$. The typical soft, blackbody-like spectra of MSPs pose a challenge for observations with {\it Chandra} due to its limited sensitivity in the soft X-ray band ($\lesssim$ 1 keV). However, a distinct subset of MSPs, known as spider pulsars, exhibit non-thermal X-ray emission produced by intra-binary relativistic shocks as a result of collisions between the pulsar wind and a matter outflow from the companion. 
Two MSPs (Seq.9 in Terzan 5 and Seq.5 in M\,28) in our sample are verified as spider pulsars,
while Seq.10 in Terzan 5 remains as a probable candidate of this class, which exhibit non-thermal spectra and substantial X-ray variations throughout the orbit cycle \citep{2022MNRAS.511.5964Z}.

A total of eleven periodic CV candidates are detected in these two GCs.
Both Terzan 5 and M\,28 are heavily obscured clusters situated in the inner Galaxy.
This raises the concern of potential interlopers, which are due to faint X-ray sources within the Galactic bulge \citep{2006ApJ...651.1098H}.
We estimate the number of potential interlopers as follows.

\begin{figure}
	\includegraphics[scale=0.335]{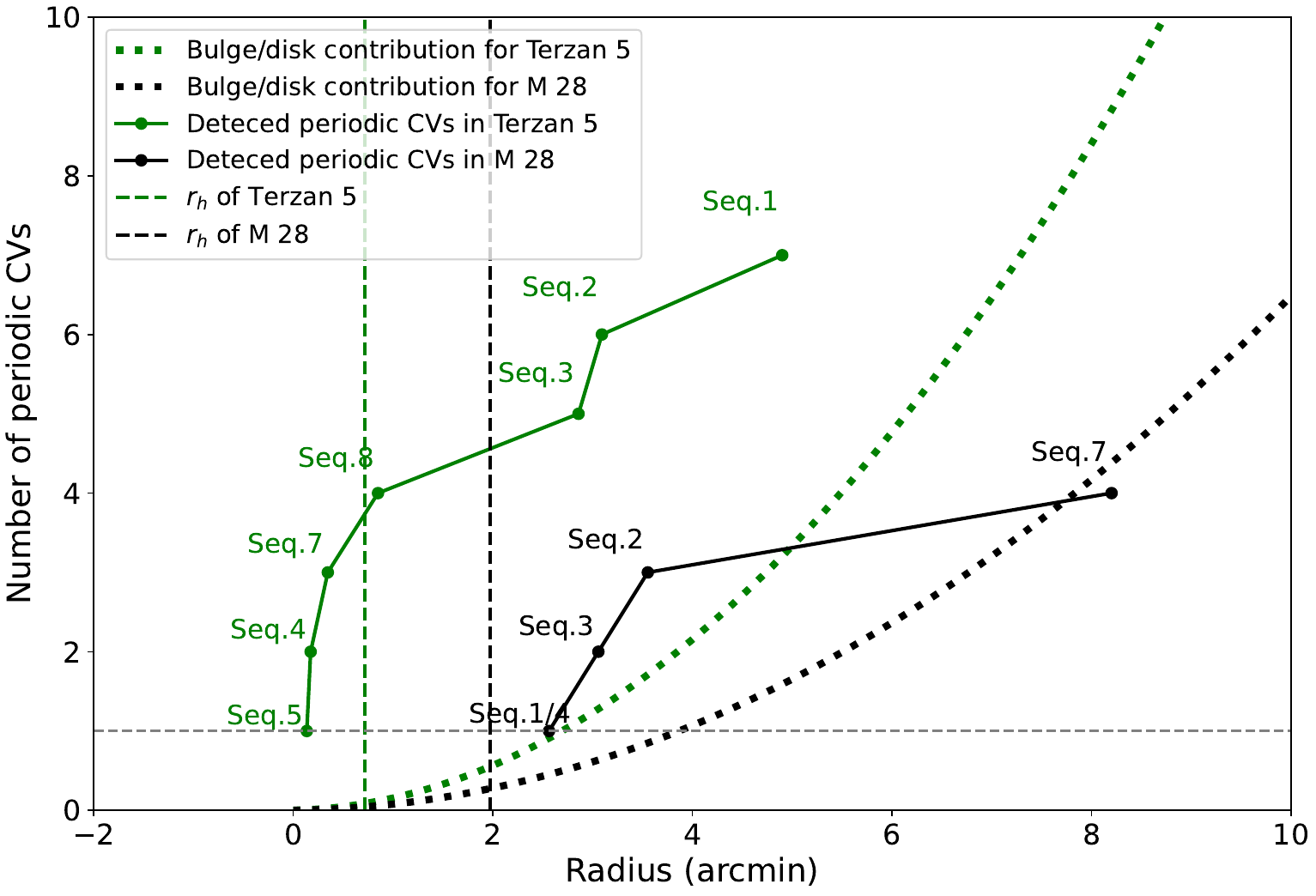}
	\caption{The green and black dotted line represents the estimated cumulative number of periodic CVs from the Galactic bulge, as a function of projected radius from the center of Terzan 5 and M 28, respectively.
	The observed cumulative distribution in Terzan 5 and M 28 is plotted by a green and black point-fold line, respectively, with the source radial position marked by its sequence number as in Table \ref{tab:srcinfo}. The half-light radius ($r_{\rm h}$) of Terzan 5 and M\,28 are also plotted as vertical dashed line.}
    \label{fig:CV_bkg}
\end{figure}

Based on the 1 Ms {\it Chandra} observations toward the LW and the same methodology as employed in the present work, \citet{2020MNRAS.498.3513B} have found 20 periodic CVs (i.e., those plotted in Figure~\ref{fig:N_P}). Assuming that this sample is representative of the bulge CV population and that the number of this population is proportional to the underlying stellar mass (i.e., a constant abundance), one may readily predict the number of bulge CVs falling into the Chandra field of Terzan 5 and M\,28. 
We utilize the three-dimensional stellar mass model of the Galaxy by \citet{2010A&A...515A..49R} to derive the stellar mass surface density in the direction of the LW, Terzan 5 and M\,28, which 
has a value of $4.6\times 10^4\rm~M_{\odot}~arcmin^{-2}$, $1.9\times 10^4\rm~M_{\odot}~arcmin ^{-2}$, and $9.6\times 10^3\rm~M_{\odot}~arcmin ^{-2}$, respectively.
The assumed scaling with mass then predicts the number of periodic CVs as a function of projected radius from the cluster center, as shown in Figure \ref{fig:CV_bkg} (green dotted line for Terzan 5 and black dotted line for M\,28). 
The observed periodic CVs are plotted for comparison, sorted by their proximity to the cluster center and labeled accordingly.

Regarding Terzan 5, the four CVs located either within or near the high-light radius $r_{\rm h}$ are highly likely to be true cluster members.
However, the three CVs situated in the outer region are more indicative of sources originating from the Galactic bulge or disk. 
Our estimation further supports this notion, as approximately three periodic CVs are expected to be identified from the Galactic bulge or disk at the projected radius of the outermost source (Seq.1).
In the case of M\,28, all CVs are situated outside the half-light radius. Specifically, three of them (Seq.1/4, Seq.3, Seq.2) are located within the range of 2 to 4 arcmin, at which point one background CV is expected from the bulge/disk. Therefore, among these three sources, at least one could be a non-member of M\,28.
The outermost source, Seq.7, positioned beyond 8 arcmin, is also quite likely a non-member.

Consequently, it is plausible that a substantial fraction of periodic CVs detected in M\,28 and Terzan 5 are unrelated to the cluster, since we cannot rule out the possibility that they originate from the Galactic bulge/disk.

\subsubsection{47 Tuc}
Detailed discussions about the 47 Tuc CV population were presented by \citet{2023MNRAS.521.4257B}. Here we highlight the most significant points.
Among the seven GCs, 47 Tuc hosts all detected periodic CVs falling in the period gap as well as a group of relatively faint CVs with unusually long orbital periods ($P\rm_{orb} \gtrsim 12\rm~h$). 
The overabundance of long-period CVs with a subgiant donor offers a compelling evidence for the recent formation of a subgroup of CVs through dynamical interactions.  
Moreover, the steep radial distribution of these periodic CVs, compared to other X-ray sources (mostly CVs and qLMXBs) in 47 Tuc, serves as additional support for their recent formation within the cluster core as a result of dynamical interactions \citep{2023MNRAS.521.4257B}.
Regarding the CVs found within the period gap, whether they possess magnetic WDs or not, their detection exclusively in 47 Tuc might be attributed to factors such as the greater mass of 47 Tuc, extended duration of observations, and relatively lower levels of extinction compared to Terzan 5 and M28. 
As indicated in Table \ref{tab:obsinfo}, 47 Tuc exhibits the highest number of bright X-ray point sources.

\begin{figure*}
	\includegraphics[scale=0.65]{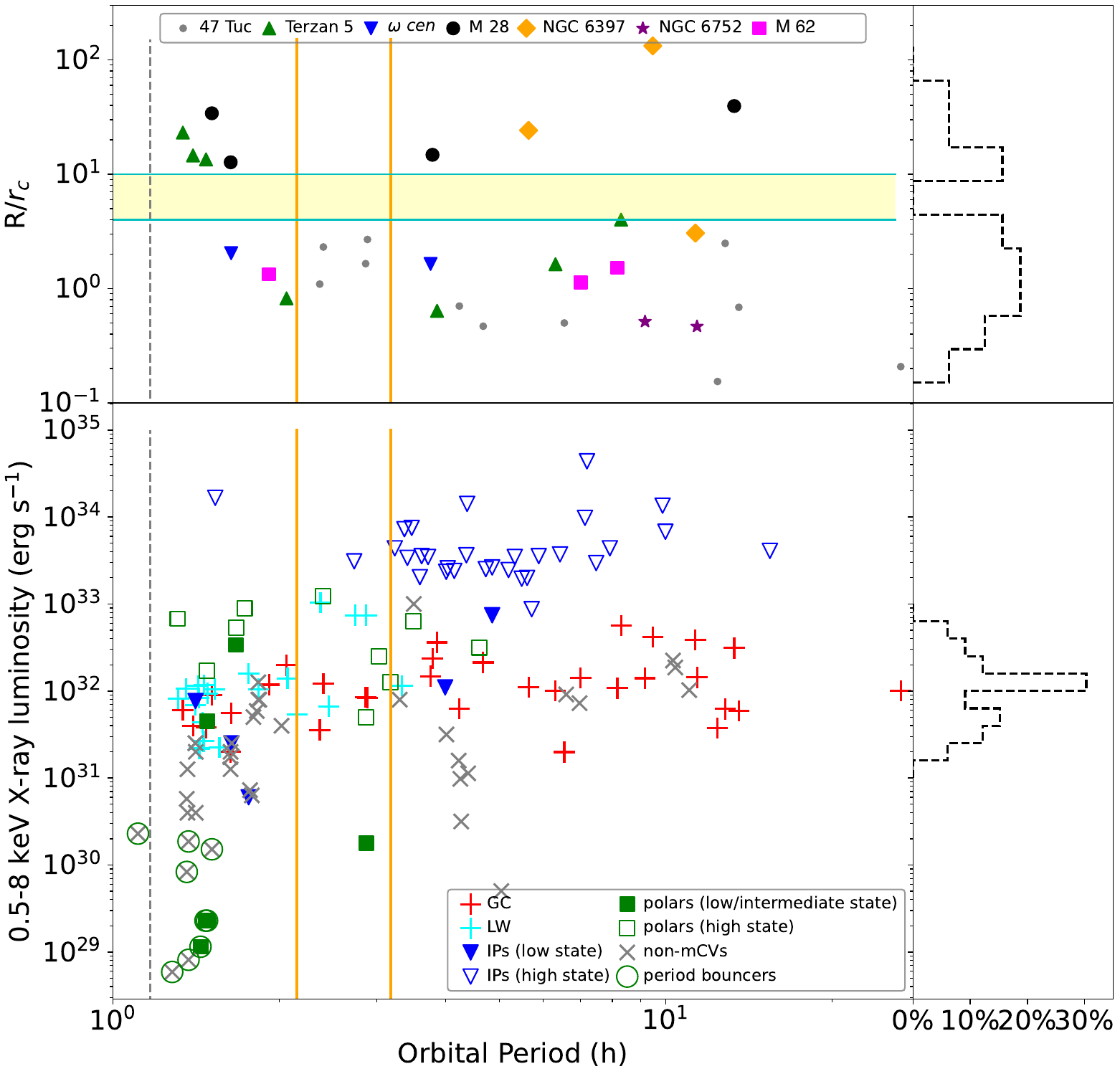}
	\caption{{\it Upper left panel}: The projected distance from the cluster center, normalized to the host GC core radius, versus the putative orbital period of the CVs identified in the seven GCs. 
CVs belonging to different GCs are labeled with different colored symbols. The yellow strip marks an apparent gap in the radial range between 4 to 10 times the core radius.
{\it Low left panel}: 
The 0.5--8 keV X-ray luminosity versus orbital period for the GC CVs (grey crosses). LW CVs (cyan `+') with an identified orbital period are plotted for comparison, as well as  different types of field CVs (polars, IPs, and non-magnetic CVs) represented by colored symbols as denoted in the insert. 
In both panels, the CV period gap and period minimum, with values taken from \citet{2011ApJS..194...28K}, are denoted by the vertical orange and grey lines, respectively.
{\it Right panels}: Histogram of the projected distance and X-ray luminosity for the GC CVs.\\
Refs for polars: \citet{2013MNRAS.432..570P,2020MNRAS.498.3457S,2023A&A...676A...7M,2021A&A...646A.181S,2021A&A...645A..56B,2023arXiv230605576O}.
Refs for IPs: 
\citet{2013MNRAS.432..570P,2019MNRAS.482.3622S,2020ApJ...898L..40L}. 
Refs for non-mCVs: 
\citet{2011MNRAS.414L..85U,2012MNRAS.419.1442P,2015MNRAS.448.3455B,2018A&A...619A..62S,2020AdSpR..66.1139N,2021MNRAS.501.2790B,2022ApJ...934..142S}
}
\label{fig:CV_profile}
\end{figure*}

\subsection{The global properties of GC CVs}

\subsubsection{Orbital distribution of CVs in different environments}
Now we turn to the global properties of GC CVs.
As shown in Figure \ref{fig:N_P}, compared with CVs in the LW and the solar neighborhood, the GC CVs show a marked difference in their orbital distribution.
The value of period gap and period minimum of CVs in Figure \ref{fig:N_P} are obtained from \citet{2011ApJS..194...28K}. However, it is noteworthy that these values are not fixed and dependent on the metallicity. Therefore, the range of the period gap exhibits intrinsic scatter rather than the simplified version presented here.  

Evidently, there is a higher proportion ($63\%$, 20 of 32) of long-period CVs (P $\gtrsim$ 3 h) compared to the local CV sample ($22\%$, 72 of 326). 
As discussed in Section \ref{subsubsec:outerCV}, it is possible that some sources in this sample are not true GC members. However, even when considering a more {\it genuine} sample, specifically the CV candidates found within the half-light radius, the proportion of long-period CVs ($70\%$, 16 of 23) remains significantly higher than that of LW.
The most natural and straightforward explanation is the selection effect, as CVs with longer periods tend to have higher X-ray luminosities, increasing the likelihood of detecting their periods within our study.
However, we emphasize that LW CV sample was constructed using the same method and procedure as for the GC CV sample. The LW field in fact provides a lower detection limit for the X-ray sources than most of the GC fields, but exhibits a paucity of long-period CVs, with 75\% of them (15 out of 20) below the period gap. 

It is also noteworthy that the sensitivity for detecting periodic sources in 47 Tuc, Terzan 5, and M\,28 is actually comparable to that for the LW, due to their closer proximity and relatively extended exposure times.
Indeed, the 1--8 keV X-ray luminosity of the faintest periodic source observed in 47 Tuc, M,28, and Terzan 5 is approximately $10^{31}\rm~erg~s^{-1}$, $10^{31}\rm~erg~s^{-1}$ and $3 \times 10^{31}\rm~erg~s^{-1}$, which is similar to the faintest source observed in the LW, with a luminosity of $10^{31}\rm~erg~s^{-1}$.
Furthermore, we have identified five short-period CVs in the observations of Terzan 5 and M\,28 with X-ray luminosities comparable to LW CVs, demonstrating our capability to detect fainter CVs with short periods.
Therefore, we can reasonably deduce that the paucity of short-period CVs in GCs cannot be solely attributed to selection effect.

There are two intrinsic factors in GC CVs that may be at least partially responsible for the significant disparity:
(i) The dynamical channel of CV formation in GCs results in more massive and younger systems with wider orbits. 
Since our sample is subject to a selection effect related to the brightness of the CVs \citep{2023MNRAS.521.4257B}, we have a higher chance to detect these dynamically-formed CVs which are expected to consist of massive, younger systems;
(ii) The binary–single or binary–binary interactions within GCs, which would lead to both the disruption of soft primordial binaries and the formation of close binaries \citep{1975AJ.....80..809H,1975MNRAS.173..729H,1993ApJ...403..256H}, suppress the formation pathway of CVs through primordial binary evolution, resulting in a lower proportion of short-period CVs in the present-day population.
The under-abundance of weak X-ray sources (mostly CVs and ABs) in GCs, compared to the Solar neighborhood and dwarf elliptical galaxies in the Local Group, as noted by \citet{2000ASPC..198..421V,2001A&A...368..137V,2015ApJ...812..130G}, further supports the notion that stellar interactions suppress the pathway for CV formation through primordial binary evolution.

\subsubsection{The radial distribution of GC CVs}
\label{sec:radial}
The upper panel of Figure \ref{fig:CV_profile} further depicts the CV orbital period versus the projected radius (normalized to the core radius) from the GC center. 
Notably, the radial profile exhibits an apparent gap between 4--10 times the core radius (marked by a yellow strip).
The appearance of this gap may be attributed to the distinct group of periodic CVs in the cluster outskirt ($R/R\rm_c > 10$).
We note that these outer CVs (excluding the most potential background source Seq.7 of M\,28) all have a high confidence in their orbital period (see the beginning of Section \ref{sec:discuss}). 
While it is likely that a significant portion of these CVs may originate from the Galactic bulge/disk rather than the GCs,   
we consider the possibility that a subset of them could be dynamically formed CVs in the GCs.

Specifically, if these CVs are of a primordial origin, their progenitors should have formed a long time ago, i.e., within $\sim$ 2 Gyr after the formation of the parent cluster \citep{2019MNRAS.483..315B}.
Taking into account the fact that the relaxation time at the half-light radius of Terzan 5, M 28 and NGC 6397 are much shorter ($\lesssim 1-2$ Gyr) \citep[][2010 ed.]{1996AJ....112.1487H}, these pre-CVs should have been able to sink to the central region, or at the very least, to within the half-light radius, contrary to the observed locations of the periodic CVs. 
Therefore, it appears more plausible that these CVs formed only recently as a result of dynamical interactions. 
In particular, dynamical interactions initiate the formation of a detached binary consisting of a WD and a main-sequence star. Since the formation process can be energetic, the resultant pre-CVs will likely be expelled far from the core \citep{2019MNRAS.483..315B}. Over a certain period, which depends on factors such as the rate of AML, the mass of the main-sequence star, and the orbital period, this detached binary eventually evolves into a CV at the cluster outskirt. 

\subsubsection{The magnetic nature of GC CVs}
The magnetic nature of CVs within GCs has long been a subject of debate, primarily due to the infrequent detection of dwarf nova (DN) outbursts \citep{1994ApJ...429..767S,1996ApJ...471..804S,2008MNRAS.388.1111P}. 
Further observational evidence supported the idea of most GC CVs being magnetic, including the identification of strong He II emission lines from certain CVs in NGC 6397 \citep{1995ApJ...455L..47G} and the much higher X-ray-to-optical ratios of GC CVs than those found among non-magnetic CVs in the field \citep{1997A&A...327..602V}.
However, these assertions have been thoroughly examined and contested by \citet{2003ApJ...596.1197E, 2006ApJ...640..288D, 2012MmSAI..83..549K, 2021gacv.workE..13B}. 
They argued that there is no direct observational evidence to conclusively establish that the majority of GC CVs are magnetic.
Moreover, one should bear in mind that our information about GC CVs is even more biased towards luminous sources than the field population.

In the present study, we construct a CV sample based solely on their periodicity, but this is still not fully immune to selection bias toward luminous systems.
In the lower panel of Figure \ref{fig:CV_profile}, the X-ray luminosity ($L\rm_X$) of the periodic GC CVs is plotted as a function of normalized projected distance. 
The $L\rm_X$ of these sources ranges between $10^{31-33}\rm~erg~s^{-1}$, with the lower bound primarily due to the typical sensitivity of the Chandra data.
Different types of field CVs with known X-ray luminosity and orbital period in the literature are also plotted for comparison.
The $L\rm_X$ of the field sources spans a much wider range, from $\sim10^{29}\rm~erg~s^{-1}$ (period bouncers) to $\gtrsim10^{34}\rm~erg~s^{-1}$ (IPs in a high state).

Within the group of GC CVs, there are certain CVs that have been confirmed to possess magnetic WDs. 
Two IPs are identified in M\,28 and 47 Tuc due to the detection of a dual period.
In addition, we have also observed some short-period CVs in M\,28 (Seq.2) and Terzan 5 (Seq.1-Seq.3), which exhibit a ``two pole'' pattern in the phase-folded light curve, indicating that they are most likely polars.
Therefore, we can reasonably infer that those CVs lying within and below the period gap are dominated by polars, with a few possibly being low-state IPs or DNe, as also indicated by their luminosity range (Figure \ref{fig:CV_profile}).

On the other hand, systems with $P_{\rm orb} \gtrsim 5$ h are more likely non-magnetic CVs, for several reasons. 
First, these systems
have similar $L_{\rm X}$ to that of field non-magnetic CVs.
Second, polars are very rare for such long orbital periods due to the requirement of an unusually strong magnetic field.  
Third, these systems are intrinsically fainter than IPs in the high state. 
Although it is possible that some of them are high luminosity IPs in the low state.
, those cases are relatively rare (e.g. Table 1 in \citealp{2023MNRAS.523.3192M}). 
For the several systems above the gap but with an orbital period $\lesssim 5$ h, they are significantly more luminous than field non-magnetic CVs. They appear to be more consistent with polars, as nearby polars are mostly found below, inside, or just above the gap \citep{2020MNRAS.498.3513B}.

In addition, it is evident that our GC CV sample lacks luminous CVs (i.e., $L_{\rm X} \gtrsim 10^{33}~\text{erg~s}^{-1}$).
Such sources, if existed, would have a low probability of being missed by our period searching process. 
Luminous CVs are typically IPs in a high state. A recent survey of X-ray sources in 38 Galactic GCs also highlights a significant under-abundance of bright IPs compared to the Galactic field \citep{2020ApJ...901...57B}. 
Therefore, there appears growing evidence that GCs in fact do not harbor a substantial population of IPs, which is contrary to previous suggestions \citep{1995ApJ...455L..47G,2006ApJ...640..288D,2006MNRAS.372.1043I}.

\section{Summary} 
\label{sec:summary}
We have conducted a systematic search for periodic X-ray signals from a large number of discrete X-ray sources in ten Galactic GCs using archival {\it Chandra} observations. 
The main findings of our study are as follows:
\begin{itemize}

\item By employing the Gregory-Loredo algorithm, we have identified 28 periodic signals originating from 27 distinct X-ray sources in 6 GCs, among which 21 are newly discovered. 
The remaining 4 GCs exhibit no period X-ray source, which is mainly due to a relatively lower sensitivity of the data.
Through analysis of their X-ray temporal and spectral properties, complemented by available optical/UV/radio information, we tentatively classify the 28 sources into 21 CVs, 5 MSPs, and 1 AB.

\item The new sample of periodic CVs from this study, combined with the 11 periodic CVs in 47 Tuc, offer the most comprehensive collection of GC CVs (32 CVs) with a probable orbital periods to date. This compilation includes the identification of eight newly discovered short-period CVs.

\item Nine CVs are found situated beyond the half-light radius of their host GCs. Although a significant portion of them may be foreground or background field CVs, some of them may be true cluster members that are recently dynamically formed and subsequently ejected to the cluster outskirt.

\item The distinctive orbital distribution exhibits a higher proportion of long-period CVs in GCs than those in the Galactic bulge and solar neighborhood. 
This disparity can be attributed to selection bias favoring younger, dynamically-formed systems and to the suppression of the pathway for CV formation through primordial binary evolution by dynamical interactions.

\item Based on their temporal/spectral properties, a considerable proportion of the GC CVs, mostly with an orbital period below or inside the period gap, are highly likely magnetic CVs. 
On the other hand, there appears a deficiency of luminous IPs ($L_{\rm X} \gtrsim 10^{33}~\text{erg~s}^{-1}$). CVs with an orbital period longer than 5 hour are more consistent with non-magnetic CVs. 
\end{itemize}

Our study highlights how high-resolution, high-sensitivity X-ray observations are promising to probe the exotic close binary populations in Galactic GCs. 

\section*{Acknowledgements}

This work is supported by the National Natural Science Foundation of China (grant 12225302). B.T. acknowledges support by the Postgraduate Research \& Practice Innovation Program of Jiangsu Province (KYCX22.0106). Z.C. acknowledges support by the National Natural Science Foundation of China (grant 12003017). D.B. acknowledges financial support from {FONDECYT} grant number {3220167}. The authors wish to thank  Jae sub Hong for valuable discussions on the GL probability. 
The authors also wish to thank the referee, Dr. Craig Heinke, for useful comments that help significantly improve the manuscript.

\section*{Data Availability}
The data underlying this article will be shared on reasonable request to the corresponding author. 
The Chandra data used in this article are available in the Chandra Data Archive (\href{https://cxc.harvard.edu/cda/}{https://cxc.harvard.edu/cda/}) by searching the Observation ID listed in Table \ref{tab:obsinfo} and in the Search and Retrieval interface, ChaSeR (\href{https://cda.harvard.edu/chaser/}{https://cda.harvard.edu/chaser/}).



\bibliography{example}{}
\bibliographystyle{mnras}



\appendix

\section{Potential false detection due to red noise}
\label{sec:confidence}
By definition, the GL probability ($P_{\rm GL}$) for a periodic signal describes the optimal data modeling and its corresponding ``goodness-of-fit'', which measures the level of the observed data originated from periodic variation, by evaluating the ``odds ratio'' between a periodic model and a constant model \citep{1992ApJ...398..146G}. 
Thus, 
$P_{\rm GL}$ stands for the likelihood that {\it ``the signal is periodic rather than constant''}, but not the likelihood that {\it ``the variability originates from an intrinsic periodic signal rather than an aperiodic one''}. 
In practice, the GL algorithm could report a substantial value of $P_{\rm GL}$ for a light curve with an aperiodic variation.

Accretion-powered systems, such as CVs and LMXBs, are known to exhibit aperiodic variability across a wide range of time scales. The so-called red noise, which is a significant component of the aperiodic variation, can potentially introduce false periodic signals, particularly at lower frequencies \citep{1989IBVS.3383....1W}. 
It is therefore instructive to estimate the possibility of false alarms among the reported signals by the GL algorithm.  

Following \citet{2023MNRAS.521.4257B},
we adopt a power-law model to describe the source power spectrum in order to account for the presence of red noise:

\begin{equation}
P(\nu)=N \nu^{-\alpha}+C_{\rm p}.
\label{eqn:LMXBps}
\end{equation}

Here $N$ is the normalization factor, $\alpha$ is the spectral index, and $C_{\rm P}$ represents the Poisson noise dictated by the mean photon flux of the source.
To mitigate the potential effect of interrupted observations in the Fourier analysis, we utilize the longest {\it Chandra} observation for each periodic source to characterize their power spectrum. 
This approach ensures a continuous and extended observation period, avoiding the influence of interruptions on the analysis.
The power spectrum of a given source is fitted with Eq.~\ref{eqn:LMXBps} using the Markov Chain Monte Carlo approach (with the python \emph{emcee} package, \citealp{2013PASP..125..306F}) to determine the best-fit parameters and errors. 

It turns out that for the majority of sources, the normalization factor cannot be reliably constrained. This is primarily attributed to the faintness of most sources, resulting in the dominance of Poisson noise.
In such cases, we opt to adopt a pure Poisson noise model instead of a power-law model.

\begin{figure*}
\centering
\includegraphics[scale=0.7]{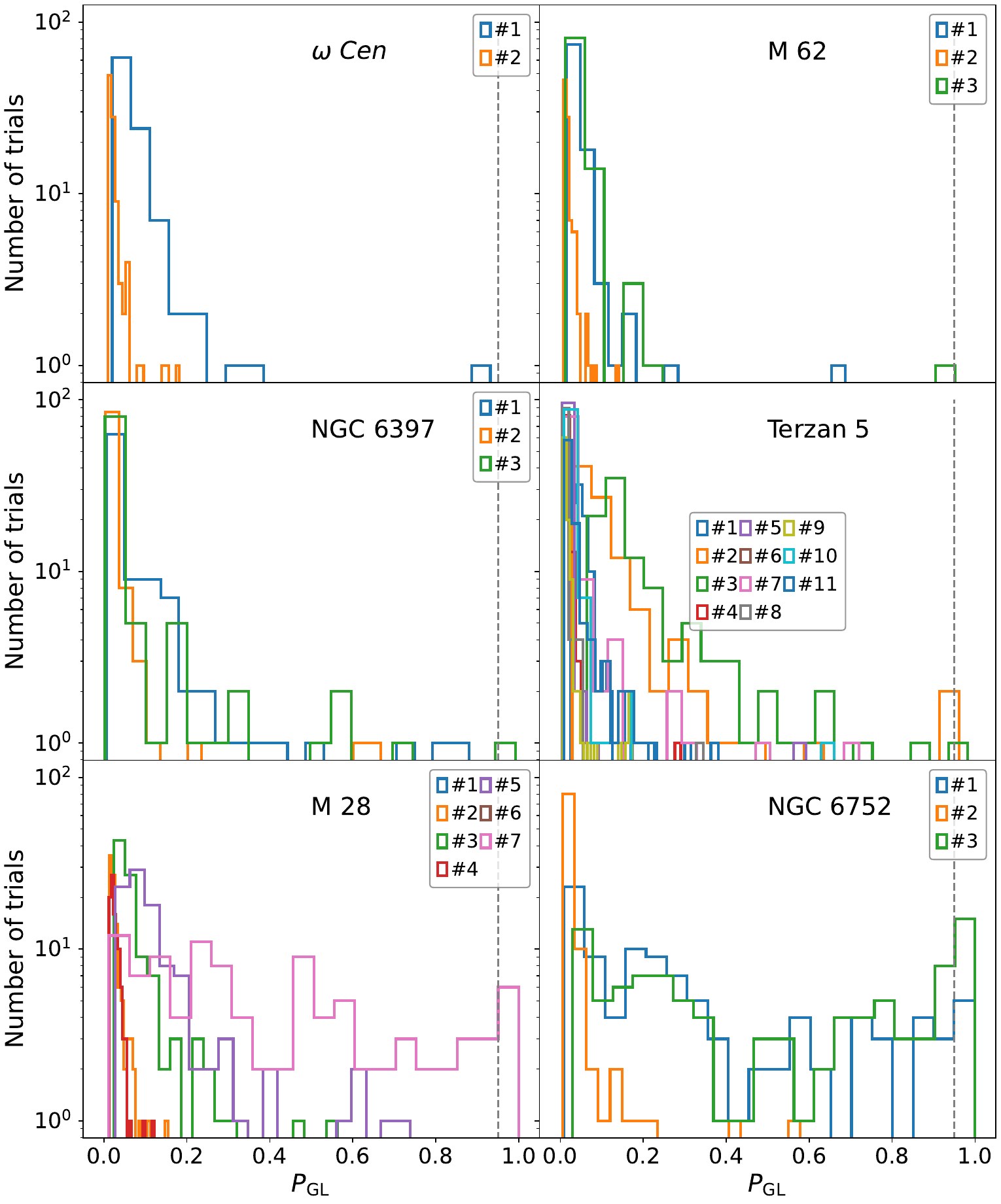}

\caption{The distribution of $P_{\rm GL}$ from 100 simulated light curve based on the modeled red noise of each periodic source (colored histograms). The grey dashed line marks $P_{\rm GL}=0.96$, the minimum value reported by the GL algorithm from the actual data. Sources in different GCs are presented in different panels.}
\label{fig:sim_hist}

\end{figure*}

Next, following the procedure proposed by \citet{1995A&A...300..707T}, we simulate 100 time series using the best-fit power spectrum model of each source, which are fed to the GL algorithm. 
The histogram of the resultant $P_{\rm GL}$ for each source is shown in Figure \ref{fig:sim_hist}.
Setting $P_{\rm GL}=96\%$ (the lowest $P_{\rm GL}$ value among our periodic signals; Table~\ref{tab:srcinfo}) as the detection threshold, the fraction of $P_{\rm GL}$ above this threshold thus represents the false alarm probability (FAP) of the source.  
As a result, all but three sources (Seq.7 of M28, Seq.1 and Seq.3 of NGC 6752) have their $P_{\rm GL}$ below 96\% in any of the 100 simulated light curves.
In other words, only three of the candidate periodic sources have an FAP greater than 1\%.

One may suspect that the period searching process would result in more false detections as the sample size increases, since it brings more trials. For instance, if a 90\% confidence level threshold is adopted, it would yield 10 false detections out of 100 sources.
 However, this holds true only when the ``optimization indicator'', such as $P_{\rm GL}$, follows the same statistics for each item in the sample.

In our study, despite the large sample size, i.e., a total of more than 2000 X-ray sources in all ten GCs for period searching, this heterogeneous sample is not suitable for determining a {\it global} confidence level threshold.
Actually, each source exhibits unique conditions, including factors such as detector response, number of counts, and background noise. The level of randomness varies significantly among different sources. 
This is clearly illustrated in Figure \ref{fig:sim_hist}. Our criterion for assessing the periodicity, i.e., the value of $P_{\rm GL}$, exhibits markedly different distributions among different sources, especially as their variability amplitudes differ. 
Therefore, the increased number of trials, attributed to the quantity of sources, does not compromise the confidence level of these signals.

\section{Inter-observation light curves of dubious periodic signals}
\label{sec:longlc}
In Appendix \ref{sec:confidence}, it is found that three of the periodic signals are subject to a substantial false alarm probability due to red noise. However, the presence of red noise does not necessarily exclude the possibility of detecting true periodic variations. 
For instance,  \citet{2023MNRAS.521.4257B} demonstrate that the GL algorithm can still uncover periodic signals in the presence of substantial red noise.  

\begin{figure}
\centering
\includegraphics[scale=0.34]{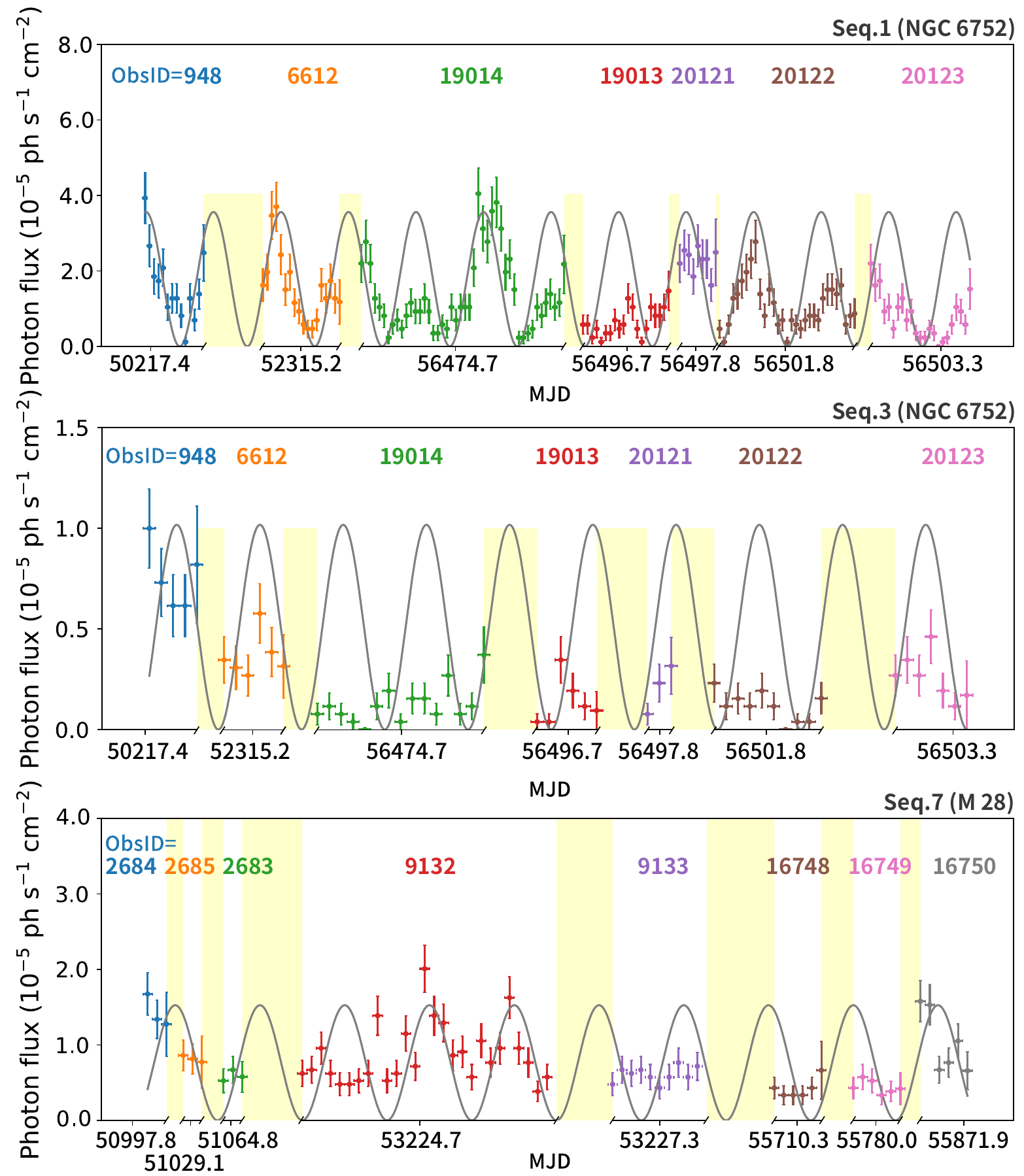}

\caption{The inter-observation light curve of Seq.1 of NGC 6752 (upper panel), Seq.3 of NGC 6752 (middle panel), and Seq.1 of M28 (lower panel) in the 0.5–8 keV energy range. The light curve includes all observations that collectively demonstrate the presence of periodic variation, with a sinusoidal curve overlaid to aid visualization.
Each set of colored data points represents a single observation, labeled by the ObsID. Additionally, yellow strips are used to denote the eliminated gaps between consecutive observations, which are integer multiples of the detected period.
}
\label{fig:longobs3}

\end{figure}

We construct the inter-observation light curves for the three sources by eliminating the observation gaps with an integer multiplication of the detected period, which makes the light curve look like a semi-continuous one, as shown in Figure \ref{fig:longobs3}. 
In the case of Seq.1 of NGC\,6752, despite the presence of significant aperiodic variability, the inter-observation light curve displays clear peak-to-peak variation following the identified period. This strongly suggests that the period is intrinsic rather than an artifact caused by red noise.

However, caution must be exercised regarding the other two sources: Seq.3 of NGC 6752 and Seq.6 of M 28. The periodic variability in these sources is not as pronounced as in Seq.1 of NGC\,6752, indicating the possibility of a spurious signal. Therefore, further observations are required to confirm the presence of periodic signals in these cases.

\bsp	
\label{lastpage}
\end{document}